\documentclass[twocolumn]{IEEEtran}
\usepackage{amsmath}
\usepackage{latexsym}
\usepackage{amssymb}
\usepackage{bm}
\newtheorem{thm}    {Theorem}
\newtheorem{lem}     {Lemma}
\newtheorem{rem}     {Remark}
\def\argmax{\mathop{\rm argmax}}

\newcommand{\defeq}{\stackrel{\rm def}{=}}

\def\cY{{\cal Y}}
\def\cZ{{\cal Z}}
\def\cN{{\cal N}}
\def\cM{{\cal M}}
\def\cD{{\cal D}}
\def\cU{{\cal U}}
\def\rE{{\rm E}}
\def\rP{{\rm P}}

\newcommand{\cP}{{\cal P}}

\newcommand{\cX}{{\cal X}}

\newcommand{\bW}{{\bf W}}
\newcommand{\bQ}{{\bf Q}}

\newcommand{\bp}{{\bf p}}

\makeatletter
\newcommand{\lleq}{\mathrel{\mathpalette\gl@align<}}
\newcommand{\ggeq}{\mathrel{\mathpalette\gl@align>}}
\newcommand{\gl@align}[2]{
\vbox{\baselineskip\z@skip\lineskip\z@
\ialign{$\m@th#1\hfil##\hfil$\crcr#2\crcr{}_{{}_{(=)}}\crcr}}}
\makeatother

\def\Label#1{\label{#1}\ [\ #1\ ]\ }
\def\Label{\label}

\begin{document}
\title{General non-asymptotic and asymptotic 
formulas \\in channel resolvability and identification capacity\\
and their application to wire-tap channel}

\author{
Masahito Hayashi
\thanks{
M. Hayashi is with Quantum Computation and Information Project, ERATO, JST,
5-28-3, Hongo, Bunkyo-ku, Tokyo, 113-0033, Japan.
(e-mail: masahito@qci.jst.go.jp)
The material in this paper was presented in part
at 2004 International Symposium on Information Theory and 
its Applications, Parma, Italy, October 2004.}}
\date{}
\maketitle
\begin{abstract}
Several non-asymptotic formulas 
are established in channel resolvability and 
identification capacity,
and they are applied to wire-tap channel.
By using these formulas, 
the $\epsilon$ capacities of
the above three problems are
considered in the 
most general setting, where 
no structural assumptions 
such as the stationary memoryless property are made on 
a channel.  
As a result, we solve
an open problem proposed in Han \& Verd\'{u}\cite{Han-V} and
Han \cite{Han-book}. 
Moreover, 
we obtain lower bounds of the exponents of 
error probability and the wire-tapper's information in wire-tap channel.
\end{abstract}
\section{Introduction}
\PARstart{I}{n} 1989,
Ahlswede \& Dueck \cite{Ah-D} proposed the identification 
code as a new framework for communication system using noisy channels.
However, 
the upper bound of the rate of the reliable identification 
codes was not solved in their paper.
In 1993, for analysis of the converse part of this problem,
Han \& Verd\'{u}\cite{Han-V} proposed the channel resolvability problem, 
in which 
we approximate the output distribution to a desired output distribution
by using a uniform input distribution with smaller support.
In particular, 
the capacity of this problem is defined as the rate of 
the maximal number of the size of support for 
every desired output distribution.
In order to discuss the channel resolvability problem,
they introduced the concepts of `general sequence of channels' and 
the `information spectrum method'.
They gave the relation between 
identification code and channel resolvability,
and succeeded in proving the converse part of
the capacity of identification code for the discrete memoryless channel.
In this method it is essential that the performances of 
several problems be characterized by using 
the probability distribution of 
the random variable with a form of `likelihood' function in this method. 
This insight is very useful for obtaining the overview of 
information theory\cite{Han-book}.
In particular, it gives a useful insight into quantum information theory
\cite{ON1,Nag-Hay,Hay-Nag}.
Therefore, Han \& Verd\'{u}'s paper\cite{Han-V} is undoubtably the landmark of 
information spectrum.

However, while Han \& Verd\'{u}'s paper gives the capacity of 
channel resolvability for general sequence of channels\cite{Han-V},
their proof of the converse part contains mistakes as is
recognized in section 6.3. of Han\cite{Han-book}.
They proved the achievability of
channel resolvability with the asymptotic zero error setting
for a general sequence of channels.
Concerning the converse part,
their proof is valid for the asymptotic $\epsilon$ error setting
when the general sequence of channels
has a strong converse property.
However, their proof is not valid in the general channel even in
the asymptotic zero error setting.

In this paper, we give several useful non-asymptotic formulas 
for identification code and channel resolvability,
which are divided into two parts.
One is the direct part of the identification code.
The existence of a good identification code is proved in Theorem \ref{t-1}.
This construction is much improved from Ahlswede \& Dueck's construction.
The other is the direct part of channel resolvability.
The existence of a good approximation regarding the output statistics
is proved in the two criteria, variational distance and
K-L divergence as in Theorem \ref{t-2}.
In this discussion, we derived upper bounds of 
the average of the variational distance and K-L divergence
between the output distribution of a given distribution $p$ 
and the output distribution of 
the input uniform distribution on $M$ elements of the input signal space,
when the $M$ elements
are randomly chosen with the distribution $p$ (Lemma \ref{lem-1-11}).
Combining Han \& Verd\'{u}'s relation between identification code and 
channel resolvability,
we derived 
the capacity of the channel resolvability for general sequence of channels
with the asymptotic zero error setting, 
which was conjectured by Han \& Verd\'{u}\cite{Han-V} 
((\ref{1-13-1}) and (\ref{1-13-2}) of Theorem \ref{t-3}).
This discussion is valid even though 
the strong converse property does not hold.

As another application, we give an upper bound of the capacity of the 
channel resolvability for a general sequence of channels
with the asymptotic $\epsilon$ error.
As a byproduct, we show that
there exists a sequence of codes whose second error probability goes to $0$
in any general sequence of channels,
and only the first error probability is asymptotically related to 
the probability distribution of 
the random variable with the form of `likelihood' 
((\ref{3-3-1}) and (\ref{3-3-2}) of Theorem \ref{t-3}).
We also derived
several lower bounds of exponent of channel resolvability in 
the stationary memoryless setting
with respective error criteria (Theorem \ref{t-5}).

Moreover, we apply our non-asymptotic formulas for channel resolvability to 
wire-tap channel,
in which there are two receivers {\it i.e.},
the eavesdropper and the normal receiver.
Wyner\cite{Wyner} introduced this wire-tap channel,
and proved that its capacity is greater then 
the difference between 
the normal receiver's information 
and the eavesdropper's information.
Csisz\'{a}r \& Narayan \cite{CN}
showed that
the capacity does not depend on the following two conditions 
for eavesdropper's information:
i) The eavesdropper's information must be less than $n \epsilon$ 
for given $ \epsilon \,> 0$, where $n$ is the number of
transmissions.
ii) The eavesdropper's information must go to $0$ exponentially.
However, 
there are no results giving an explicit lower 
bound of the optimal exponents of wire-tapper's information.

Indeed, this problem is closely related to the channel resolvability 
as follows.
In Wyner's proof \cite{Wyner}, 
in the asymptotic i.i.d. setting with a large enough number $M$,
he essentially showed that
when $M$ elements of the input signal space
are randomly chosen with a given distribution $p$,
the output distribution of the distribution $p$ 
can be approximated with a high probability by the output distribution of 
the input uniform distribution on the above 
$M$ elements of the input signal space.
This idea is also applied in 
Devetak\cite{Deve} and Winter, Nascimento \& Imai \cite{WNI}.
Using the same idea in the non-asymptotic setting, 
we can apply our formulas of channel resolvability 
to wire-tap channel, and
derive a good non-asymptotic formula for 
wire-tap channel (Theorem \ref{3-6}).
As consequences we obtain 
the capacity of general sequence of wire-tap channel (Theorem \ref{t-1-17}),
and lower bounds of the exponents 
of error probability and the wire-tapper's information in 
the stationary memoryless setting (Theorem \ref{t-7}).
We can expect that these results will be applied to evaluations of 
the security of channels.

Finally, we should remark that our non-asymptotic resolvability formula 
regarding 
variational distance can be regarded as 
essentially the same results as 
Oohama\cite{Oohama2}'s formula, where
he treated the partial resolvability.
Furthermore, he also derived a lower bound of exponent of 
channel resolvability by type method\cite{Oohama}.
\section{Identification code in non-asymptotic setting}
Let $W: x \mapsto W_x$ be an arbitrary channel with the input alphabet
$\cX$ and the output alphabet $\cY$.
The identification channel code for the channel $W$
is defined in the following way.
First, let $\cN =\{ 1, \ldots, N\}$ be a set of
messages to be transmitted, and
denote by $\cP(\cX)$ the set of all probability distribution
over $\cX$. A transmitter prepares $N$ probability
distributions $Q_1, \ldots, Q_N \in \cP(\cX)$.
If the transmitter wants to send a message $i \in \cN$,
an encoder generates an input sequence $x_i \in \cX$
randomly subject to the probability 
distribution $Q_i$.
In this case, the output signal $y$ obeys the distribution
$W_{Q_i}$, where
the output distribution $W_p$ of a given input distribution
$p$ is defined as
\begin{align*}
W_p(y)\defeq \sum_{x} p(x) W_x(y).
\end{align*}
On the other hand, at the decoder side an $N$-tuple of decoders
is prepared.
For every $i=1, \ldots, N$, the $i$-th decoder judges that
$i \in \cN$ is transmitted if a channel output $y$ belongs to 
$\cD_i$, where
$\{ \cD_1, \ldots, \cD_N\}$ are $N$ subsets of $\cY$ in advance.
The $i$-th decoder judges that a message different from $i \in \cN$
if $y \notin \cD_i$.
Here, $\cD_i$ is called the {\it decoding region},
of the message $i$.
It is not required that $\cD_1, \ldots, \cD_N$ be disjoint.
In the identification coding problem, the $i$-th decoder 
is only interested in transmission of the corresponding 
message $i$.
Thus, we call the tuple of 
$\Phi \defeq (N,\{ Q_1, \ldots Q_N \},\{ \cD_1, \ldots, \cD_N\})$
an identification code of channel $W$.
The performance of this code can be characterized by the following three 
quantities.
One is the size $N$ of the message sent and is denoted by $|\Phi|$, and 
the others are the maximum values of the two-type error probabilities
given as:
\begin{align*}
\mu (\Phi) \defeq \max_i  W_{Q_i} (\cD_i^c) , \quad
\lambda (\Phi) \defeq \max_{i\neq j} W_{Q_j} (\cD_i) ,
\end{align*}
where $\cD_i^c$ is the complement set of $\cD_i$.
Concerning this problem, as discussed in the following theorem,
the `likelihood' function $\frac{W_x}{W_p}(y)
\defeq \frac{W_x(y)}{W_p(y)}$ suitablely characterizes
the performance of good identification codes.
\begin{thm}\Label{t-1}
Assume that real numbers
$\alpha, \alpha',\beta,\beta',\tau, \kappa \,> 0$ 
satisfy 
\begin{align}
&\kappa \log (\frac{1}{\tau}- 1)  \,> \log 2 +1, ~
1/3 \,> \tau\,> 0,~1 \,> \kappa \,> 0,
\Label{3-3-5}\\
&1  \,> \frac{1}{\alpha} + \frac{1}{\alpha'} ,~
\gamma  \defeq 1 - \frac{1}{\beta}- \frac{1}{\beta'}\,> 0.
\Label{3-3-4}
\end{align}
Then, 
for any integer $M \,> 0$, any real number $C \,>0$,
any channel $W$,
and any probability distribution
$p \in \cP(\cX)$, 
there exists an identification code $\Phi$ such that
\begin{align*}
\mu(\Phi) &\le \alpha \beta \rE_{p,x}
W_x \left\{ y \left| \frac{W_x}{W_p}(y) \le C\right.\right\} \\
\lambda(\Phi) & \le \kappa + \alpha'\beta'\frac{1}{C}
\left\lceil \frac{M}{\gamma}\right\rceil,\quad
|\Phi| = 
\left\lfloor\frac{e^{\tau M}}{Me}\right\rfloor
\end{align*}
if
\begin{align}
\beta \rE_{p,x}
W_x \left\{ y \left| \frac{W_x}{W_p}(y) \le C\right.\right\}
+
\alpha'\beta'\frac{1}{C}
\left\lceil \frac{M}{\gamma}\right\rceil
< 1,\label{11-23-7-q}
\end{align}
where $\rE_{p,x}$ denotes the expectation
concerning random variable $x$ obeying the probability distribution $p$.
\end{thm}
In the following, we omit $x$ or $p$ in the notation $\rE_{p,x}$,
and abbreviate the set $\left\{ y \left| \frac{W_x}{W_p}(y) \le C\right.
\right\}$
as $\left\{ \frac{W_x}{W_p}(y) \le C\right\}$.
we also denote the probability
that the random variable $X$ belongs to
the set $\cD$, by $\rP_X (\cD)$ or $\rP_X \cD$.
If we do not need to take note of the random variable $X$,
we simplify it to $\rP (\cD)$ or $\rP \cD$.
This theorem is proven by using the following lemma.

\begin{lem}\Label{A-D}
(Ahlswede and Dueck\cite{Ah-D})
Let $\cM$ be an arbitrary finite set of the size $M=|\cM|$.
Choose constants $\tau$ and $\kappa$ satisfying the condition (\ref{3-3-5}).
Then there exist $N (\defeq 
\lfloor\frac{e^{\tau M}}{Me}\rfloor)$ subsets 
$A_1 , \ldots , A_N \subset \cM$
satisfying 
\begin{align}
|A_i|= \lfloor \tau M \rfloor ,\quad
|A_i \cap A_j| \,< \kappa \lfloor \tau M \rfloor (i \neq j).
\Label{3-3-7}
\end{align}
\end{lem}

\noindent
\quad{\it Proof of Theorem \ref{t-1}:}
In this proof, the subset 
$\cU_x \defeq \left\{  \frac{W_x}{W_p}(y) > C 
\right\}$ plays an important role.
First, we assume the existence of 
$M$ distinct elements $x_1, \ldots, x_M$ of $\cX$ satisfying
\begin{align}
W_{x_i} (\cU_{x_i}^c) &\le \alpha \beta \rE_p
W_x \left\{  \frac{W_x}{W_p}(y) \le C \right\} ,\Label{3-3-8}\\
W_{x_i} \left(\bigcup_{j\neq i} \cU_{x_j}\right) &\le 
\alpha'\beta'\frac{1}{C}
\left\lceil \frac{M}{\gamma}\right\rceil .\Label{3-3-9}
\end{align}
From Lemma \ref{A-D},
we can choose 
$N \defeq 
\lfloor\frac{e^{\tau M}}{Me}\rfloor$ subsets 
$A_i , \ldots, A_N$ of the set
$\{x_1, \ldots, x_M\}$ satisfying 
(\ref{3-3-7}).
Let $Q_i$ be 
the uniform distribution on the subset $A_i$ whose cardinality is
$\lfloor \tau M \rfloor $, that is,
$Q_i$ is defined as
\begin{align}
Q_i(x)\defeq \frac{1}{|A_i|} \sum_{x' \in A_i}
1_{x'}(x),
\end{align}
where $1_{x'}=1_{x'}(x)$ is an indicator function
taking value 1 if $x=x'$ and $0$ otherwise.
Defining the subset $\cD_i$ as $\cD_i\defeq \cup_{x \in A_i} \cU_{x}$,
we evaluate
\begin{align*}
& W_{Q_i}(\cD_i) =
\sum_{x \in A_i}\frac{1}{|A_i|}
W_{x} (\cD_i)
\ge
\sum_{x \in A_i}\frac{1}{|A_i|}
W_{x} (\cU_x) \\
\ge & 1- \alpha \beta \rE_p
W_x \left\{  \frac{W_x}{W_p}(y) \le C\right\},\\
& W_{Q_i}(\cD_j)
= \sum_{x \in A_i}\frac{1}{|A_i|}
W_{x} (\cD_j) \\
= &
\sum_{x \in A_i \cap A_j }\frac{1}{|A_i|}
W_{x} (\cD_j)
+
\sum_{x \in A_i \cap A_j ^c }\frac{1}{|A_i|}
W_{x} (\cD_j)\\
\le &
\frac{|A_i \cap A_j |}{|A_i|}
+
\sum_{x \in A_i \cap A_j ^c }\frac{1}{|A_i|}
W_{x} \left(\bigcup_{x'\neq x} \cU_{x'}\right)\\
\le &
\kappa
+
\alpha'\beta'\frac{1}{C}
\lceil \frac{M}{\gamma}\rceil.
\end{align*}
Therefore, we obtain the desired argument.

Next, we prove the existence of $M$ elements and 
$M$ subsets satisfying (\ref{3-3-8}) and (\ref{3-3-9})
by a random coding method.
Let $M'$ be $\lceil \frac{M}{\gamma}\rceil$,
and $X=(X_1, \ldots, X_{M'})$ be $M$ independent and 
identical random variables 
subject to the probability distribution
$p \in\cP(\cX)$,
then we have
\begin{align*}
W_p(\cU_x) \le \frac{1}{C} W_x(\cU_x) \le \frac{1}{C}.
\end{align*}
Using this inequality, we obtain
\begin{align*}
& \rE_{X}
\frac{1}{M'}
\sum_{i=1}^{M'}
W_{X_i}
\left(\bigcup_{j \neq i} \cU_{X_j}\right)
\le 
\rE_{X}
\sum_{j=1}^{M'}
\frac{1}{M'}
\sum_{i\neq j}
W_{X_i}(\cU_{X_j}) \\
&=
\sum_{j=1}^{M'}
\rE_{X_j}
\frac{M'-1}{M'}
W_p(\cU_{X_j})
\le
\sum_{j=1}^{M'}
\rE_{X_j}
\frac{M'-1}{M'C}
\le \frac{M'-1}{C}.
\end{align*}
Further,
\begin{align*}
&\rE_{X}\frac{1}{M'}\sum_{i=1}^{M'}
W_{X_i}(\cU_{X_i}^c)
=
\sum_{i=1}^{M'}
\rE_{X_i}W_{X_i}\frac{(\cU_{X_i}^c)}{M'}
= \rE_p W_x(\cU_x^c) .
\end{align*}
Using the Markov inequality
$\rP_X \{ X  \,> \alpha \rE X\}\,< \frac{1}{\alpha}$,
{\em i.e.}, 
$\rP_X \{ X  \le \alpha \rE X\}\,> 1-  \frac{1}{\alpha}$,
we can show that
\begin{align*}
\rP_X
\left\{
\frac{1}{M'}
\sum_{i=1}^{M'}
W_{X_i}
\left(\bigcup_{j \neq i} \cU_{X_j}\right)
\le
\alpha' \frac{M'-1}{C}
\right\}
& \,>
1- \frac{1}{\alpha'} \\
\rP_X
\left\{
\frac{1}{M'}\sum_{i=1}^{M'}
W_{X_i}(\cU_{X_i}^c)
\le
\alpha \rE_p W_x(\cU_x^c) 
\right\}
& \,>
1- \frac{1}{\alpha}.
\end{align*}
Since
$(1- \frac{1}{\alpha'})+(1- \frac{1}{\alpha})\,>1$,
there exist $M$ elements
$x_1, \ldots, x_{M'}$ such that
\begin{align*}
\frac{1}{M'}
\sum_{i=1}^{M'}
W_{x_i}
\left(\bigcup_{j \neq i} \cU_{x_j}\right)
& \le
\alpha' \frac{M'-1}{C} \\
\frac{1}{M'}\sum_{i=1}^{M'}
W_{x_i}(\cU_{x_i}^c)
& \le
\alpha \rE_p W_c(\cU_x^c) .
\end{align*}
In the following, 
the above $M$ elements $x_1, \ldots, x_{M'}$ are fixed,
and we only focus on 
the random variable $i$ subject
to the uniform distribution on the set $\{1, \ldots, M' \}$.
Combining Markov inequality and the preceding inequalities,
we have
\begin{align*}
\rP_i \left\{
W_{x_i}
\left(\bigcup_{j \neq i} \cU_{x_j}\right)
\le \beta' \alpha' \frac{M'-1}{C} 
\right\}
& \,> 1- \frac{1}{\beta'} \\
\rP_i \left\{
W_{x_i}(\cU_{x_i}^c)
\le
\beta \alpha \rE_p W_x(\cU_x^c)\right\}
& \,> 1- \frac{1}{\beta} .
\end{align*}
Hence, we obtain
\begin{align*}
& \rP_i \left\{
\begin{array}{c}
W_{x_i}
\left(\bigcup_{j \neq i} \cU_{x_j}\right)
\le \beta' \alpha' \frac{M'-1}{C} ,\\
W_{x_i}(\cU_{x_i}^c)
\le
\beta \alpha \rE_p W_x(\cU_x^c)
\end{array}
\right\} \\
\,> &(1- \frac{1}{\beta} )+
(1- \frac{1}{\beta'})-1= \gamma,
\end{align*}
which yields
\begin{align*}
\left| \left\{
i\left|
\begin{array}{c}
W_{x_i}
\left(\bigcup_{j \neq i} \cU_{x_j}\right)
\le \beta' \alpha' \frac{M'-1}{C} ,\\
W_{x_i}(\cU_{x_i}^c)
\le
\beta \alpha \rE_p W_x(\cU_x^c)
\end{array}
\right.\right\} \right|
\,>  \lfloor \gamma M'\rfloor.
\end{align*}
Since
$\lfloor \gamma M'\rfloor
= \lfloor\gamma 
\lceil \frac{M}{\gamma} \rceil \rfloor\ge M$,
there exist $M$
elements of $\cX$ satisfying (\ref{3-3-8}) and (\ref{3-3-9}).
Here, one may think that these $M$ elements 
may not be distinct.
However, if $x_i= x_{i'} (i \neq i')$,
the relation $W_{x_i}(\cU_{x_i}^c)
+ 
W_{x_i} \left(\bigcup_{j\neq i} \cU_{x_j}\right) \ge 1$
holds.
From condition (\ref{11-23-7-q}),
this contradicts 
(\ref{3-3-8}) and (\ref{3-3-9}).
Hence, we obtain the desired bound.
\endproof

\section{Channel resolvability in non-asymptotic setting}
In the channel resolvability, 
we choose 
$M$ elements $x_1, \ldots, x_M$ in the input set $\cX$ 
for every probability distribution $p \in \cP(\cX)$,
such that 
the output distribution of 
the input distribution 
\begin{align*}
\sum_{i=1}^M \frac{1}{M} 1_{x_i}
\end{align*}
close enough to 
the output distribution of 
$p$ through the channel $W$.
In particular, 
we call the distribution 
with the preceding form
an $M$-type.
In this setting, 
our purpose is to disenable
the receiver of the given channel $W$
to distinguish whether
the sender generates the input signal based on 
`the given distribution $p$'
or `the $M$-type $\sum_{i=1}^M \frac{1}{M} 1_{x_i}$ 
with a smaller number $M$'.
This kind of indistinguishability  
can not be applied to any realistic model, directly,
but it can be technically related to wire-tap channel.
In particular, 
we prove Lemma \ref{lem-1-11} in this section as
the technically essential part, but
this lemma is also the technically essential part
for the direct part of wire-tap channel.

In the following, we call
the pair of the integer $M$ and
the $M$ elements $x_1, \ldots, x_M$ of $\cX$,
a resolvability code $\Psi$ with the size $|\Psi|\defeq M$.
The performance of a resolvability code $\Psi$
is characterized by its size $|\Psi|$ and
the variational distance
\begin{align*}
\epsilon (\Psi,W_p)
\defeq d\left( 
\sum_{i=1}^M \frac{1}{M} W_{x_i}, W_p\right),
\end{align*}
where
the variational distance $d(p,q)$
defined by
\begin{align*}
d(p,q)= \sum_y |p(y)-q(y)|,
\end{align*}
which equals the $l_1$ norm $\|p-q\|_1$.
Another characterization of its performance is given by 
K-L divergence
\begin{align}
D(\Psi,W_p )\defeq D(\sum_{i=1}^M \frac{1}{M} W_{x_i}\| W_p),
\Label{3-5-1}
\end{align}
where
$D(p\|q)\defeq \sum_y p(y)\log \frac{p(y)}{q(y)}$.
\begin{thm}\Label{t-2}
For any integer $M \,> 0$, any real number $C \,> 0$ and
any probability distribution $p \in \cP(\cX)$,
there exists a resolvability code $\Psi$ such that
$|\Psi|= M$ and 
\begin{align}
\epsilon (\Psi, W_p) &\le 2 \delta_{p,W,C}
+ \sqrt{\frac{\delta_{p,W,C}'}{M}} ,
\Label{3-3-10}\\
\delta_{p,W,C}' &\le C , \nonumber
\end{align}
where $\delta_{p,W,C}
\defeq \rE_p W_x \left\{ \frac{W_x}{W_p}(y) \,> C\right\}$,
and $\delta_{p,W,C}'\defeq 
\rE_p \frac{W_x^2}{W_p} 
\left\{ \frac{W_x}{W_p}(y) \le C\right\}$.
If the cardinality $|\cY|$ is finite,
for any $0\,> t \ge -1/2$, 
there exists a resolvability code $\Psi'$ such that
$|\Psi'|=M$ and either of 
\begin{align}
D(\Psi',W_p) \le &
\frac{\log (1+M^t e^{\phi(t|W,p)})}{-t},\Label{6-26-1}\\
D(\Psi',W_p) \le & \eta(\delta_{p,W,C})+  \delta_{p,W,C}\log |\cY| + 
\frac{\delta_{p,W,C}'}{M}\Label{7-1-3}
\end{align}
holds, where 
$\eta(x)\defeq -x\log x$
and $\phi(t|W,p)\defeq \log \sum_y 
(\rE_p W_x^{1/(1+t)}(y))^{1+t}$.
\end{thm}
\begin{rem}
The partial resolvability version of inequality (\ref{3-3-10})
has been obtained by Oohama\cite{Oohama2}.
Inequality (\ref{3-3-10}) can be regarded as 
the essentially same result as Oohama's inequality.
\end{rem}
\begin{proof}
In the following,
the indicator functions 
$I_x$ and $I_x^c$ 
on the sets $\cU_x
=\left\{  \frac{W_x}{W_p}(y) > C \right\}$
and their compliment sets $\cU_x^c$
play important roles.
In our proof of Theorem \ref{t-2}, 
we use the random coding method, {\it i.e.},
we consider the $M$ independent and identical random variables 
$X=(X_1, \ldots, X_M)$
subject to $p$. 
Using the notations:
\begin{align*}
W_x^{\alpha}(y)&\defeq W_x(y)I_x^c(y) ,\quad
W_x^{\beta}(y)\defeq W_x(y)I_x(y) ,\\
W_p^{\alpha}(y)&\defeq \rE_p W_x^{\alpha}(y) ,\quad
W_p^{\beta}(y)\defeq \rE_p W_x^{\beta}(y) ,\\
W_X^{\alpha}(y)&\defeq \frac{1}{M} \sum_{i=1}^M W_{X_i}^{\alpha}(y), \quad
W_X^{\beta}(y)\defeq \frac{1}{M} \sum_{i=1}^M  W_{X_i}^{\beta}(y), \\
W_X^{M}(y)&\defeq W_X^{\alpha}(y)+ W_X^{\beta}(y)
= \frac{1}{M} \sum_{i=1}^M W_{X_i}(y),
\end{align*}
we have the following lemma.
\begin{lem}\Label{lem-1-11}
The $M$ random variables 
$X=(X_1, \ldots, X_M)$ satisfy the following inequality
\begin{align}
\rE_X
\left\| W_X^{M} - W_p \right\|_1 
\le & 2 \delta_{p,W,C}
+ \sqrt{\frac{\delta_{p,W,C}'}{M}} \Label{1-11-1}\\
\rE_X D(W_X^{M}\|W_p) 
\le &
\frac{\log (1+ M^t e^{\phi(t|W,p)})}{-t}\Label{1-11-3}
\end{align}
for $0 \,> t \ge - \frac{1}{2}$.
If the cardinality $\cY$ is finite,
the inequality 
\begin{align}
\rE_X D(W_X^{M}\|W_p) 
\le & \eta(\delta_{p,W,C})+  \delta_{p,W,C}\log |\cY| + 
\frac{\delta_{p,W,C}'}{M} \Label{1-11-2}
\end{align}
holds.
\end{lem}

Since there exists a resolvability code $\Psi$ 
with the size $M$ such that 
\begin{align*}
\epsilon (\Psi,W_p)\le \rE_X
\left\| W_X^{M} - W_p \right\|_1 ,
\end{align*}
the inequality (\ref{1-11-1}) guarantees the existence 
of a resolvability code $\Psi$ satisfying (\ref{3-3-10}).
On the other hand, 
the relation $\frac{W_{x}^{\alpha}(y)}{W_p(y)}= 
\frac{W_{x}(y)}{W_p(y)}I_{x}^c(y)\le C$ holds.
Thus,
\begin{align*}
\delta_{p,W,C}'=
\sum_y W_x(y) \frac{W_{x}^{\alpha}(y)}{W_p(y)} \le C.
\end{align*}
Similarly, 
since there exists a resolvability code $\Psi$ 
with the size $M$ such that 
\begin{align*}
D(\Psi',W_p) \le 
\rE_X D(W_X^{M}\|W_p) ,
\end{align*}
the inequalities (\ref{1-11-3}) and (\ref{1-11-2}) guarantees the existence 
of a resolvability code $\Psi$ satisfying (\ref{6-26-1}) and (\ref{7-1-3}).
\end{proof}
\quad{\it Proof of Lemma \ref{lem-1-11}}:

First, we show (\ref{1-11-1}). 
Since 
\begin{align*}
 \delta_{p,W,C}
=\rE_{p,x} W_{x} (\cU_x)
= \rE_p \| W_x^{\beta}\|
= \| W_p^{\beta} \|,
\end{align*}
we can evaluate
\begin{align*}
& \rE_X
\left\| W_X^{M} - W_p \right\|_1 \\
= &
\rE_X
\left\| W_X^{\alpha} - W_p^{\alpha}
+W_X^{\beta} - W_p^{\beta} \right\|_1 \\
\le &
\rE_X
\left\| W_X^{\alpha}- W_p^{\alpha} \right\|_1 
 + \sum_{i=1}^M \frac{1}{M}\rE_X \left\|W_{X_i}^{\beta}\right\|_1
+\left\|W_p^{\beta}\right\|_1 \\
= &
\rE_X
\left\| W_X^{\alpha}- W_p^{\alpha} \right\|_1 
+ 2\rE_{p,x} W_{x} (\cU_x).
\end{align*}
Next, we focus on 
the Schwarz inequality regarding 
the random variable
$l_X(y) \defeq \frac{W_X^{\alpha}}{W_p}(y)$
and the sign function $\tilde{l}_X(y)\defeq
\frac{l_X(y)}{|l_X(y)|}$ 
(we can check that $\tilde{l}_X^2=1$.),
then we obtain
\begin{align*}
&(\|W_p l_X\|_1 )^2= 
(\rE_{W_p} |l_X(y)|)^2= (\rE_{W_p} l_X (y)\tilde{l}_X(y))^2 \\
\le &\rE_{W_p} l_X^2(y) \rE_{W_p} \tilde{l}_X^2(y)
= \rE_{W_p} l_X^2(y) .
\end{align*}
Thus, the Jensen inequality yields that
\begin{align*}
\left( 
\rE_X
\left\| W_{X}^{\alpha} - W_p^{\alpha} \right\|_1
\right)^2
\le \rE_X
\left\| W_{X}^{\alpha} - W_p^{\alpha} \right\|_1^2
\le \rE_X \rE_{W_p} l_X^2 (y).
\end{align*}
Since $\rE_x \frac{W_x^{\alpha}(y)}{W_p(y)} = 
\frac{W_p^{\alpha}(y)}{W_p(y)}$, we have
\begin{align*}
& \rE_X \rE_{W_p} l_X^2 
=\rE_{W_p} \rE_X l_X^2(y) \\
= &
\rE_{W_p} \rE_X
\frac{1}{M^2} \sum_{i=1}^M 
\left(\frac{W_{X_i}^{\alpha}(y)}{W_p(y)}
-\frac{W_p^{\alpha}(y)}{W_p(y)}\right)^2 \\
= &
\rE_{W_p} 
\frac{1}{M} \rE_x
\left (\left(\frac{W_{x}^{\alpha}(y)}{W_p(y)}\right)^2 
-\left(\frac{W_p^{\alpha}(y)}{W_p(y)}\right)^2 \right)\\
\le &
\rE_x
\frac{1}{M} \rE_{W_p} 
\left(\frac{W_{x}^{\alpha}(y)}{W_p(y)}\right)^2 
= \frac{\delta_{p,W,C}'}{M}.
\end{align*}
Therefore, we obtain
\begin{align*}
\rE_X\left\| \sum_{i=1}^M \frac{1}{M} W_{X_i} 
- W_p \right\|_1 
\le 2 \delta_{p,W,C}
+ \sqrt{\frac{\delta_{p,W,C}'}{M}} .
\end{align*}
Hence, we obtain (\ref{1-11-1}).

Next, we show (\ref{1-11-2}).
Since $\frac{W_X^{M}(y)}{W_p(y)} 
\le \frac{1}{W_p^{\beta}(y)}$,
by using the inequality $\log x \le x-1$,
we can evaluate
\begin{align*}
&\rE_X D(W_X^{M}\|W_p) \\
=& \rE_X \sum_y \left( W_X^{\alpha}(y)\log 
\frac{W_X^{M}(y)}{W_p(y)}
+ \sum_y W_X^{\beta}(y)\log 
\frac{W_X^{M}(y)}{W_p(y)}\right) \\
\le& \rE_X \sum_y \left( W_X^{\alpha}(y)\left( 
\frac{W_X^{M}(y)}{W_p(y)}-1 \right)
+ W_X^{\beta}(y)\log 
 \frac{1}{W_p^{\beta}(y)} \right)\\
=&\sum_y 
\rE_X W_X^{\alpha}(y)
\left( \frac{W_X^{M}(y)}{W_p(y)}-1 \right)
+ \sum_y W_p^{\beta}(y)\log 
\frac{1}{W_p^{\beta}(y)} .
\end{align*}
Regarding the first term,
we can calculate
\begin{align*}
& \sum_y \rE_X W_X^{\alpha}(y)
\left( \frac{W_X^{M}(y)}{W_p(y)}-1 \right)\\
=&
\sum_y \rE_X \frac{1}{M^2}\sum_{i,j}
W_{X_i}^{\alpha}(y)
\left( \frac{W_{X_j}(y)}{W_p(y)}-1 \right)\\
= & \sum_y \frac{1}{M}\rE_{p,x}
W_{x}^{\alpha}(y)
\left( \frac{W_{x}(y)}{W_p(y)}-1 \right) \\
\le &\sum_y \frac{1}{M}\rE_{p,x}
\frac{W_{x}^{\alpha}(y)}{W_p(y)}
W_{x}(y) 
= \frac{\delta_{p,W,C}'}{M},
\end{align*}
where we use the relation $\rE_X
W_{X_i}^{\alpha}(y)
\left( \frac{W_{X_j}(y)}{W_p(y)}-1 \right)
=0$ for $i \neq j$.
Concerning the second term, 
letting $K\defeq \sum_y W_p^{\beta}(y)$,
we have 
\begin{align*}
&\sum_y W_p^{\beta}(y)\log 
\frac{1}{W_p^{\beta}(y)} \\
=&
- K \log K - K \sum_y \frac{W_p^{\beta}(y)}{K}
\log \frac{W_p^{\beta}(y)}{K}\\
\le &\eta( \sum_y W_p^{\beta}(y))
+  \sum_y W_p^{\beta}(y) \log |\cY|,
\end{align*}
because $\log |\cY|$ is the maximal entropy 
of the distribution on the probability space $\cY$.
Since
$\sum_y W_p^{\beta}(y)=
\delta_{p,W,C}$, 
we obtain 
(\ref{1-11-2}).

Finally, we prove (\ref{1-11-3})
by a different method.
The quantity $\rE_X D(W_X^{M}\|W_p)$ can be regarded as
the mutual information of channel
$X\mapsto W_X^{M}$ with the input probability
$p^M(X)$ which equals the $M$-fold i.i.d. of $p$.
We can check that 
the function $t \mapsto \phi(t|W^{M},p^M)$
satisfies the following property:
\begin{align*}
\phi(0|W^{M},p^M)&= 0 \\
\left.\frac{\,d \phi(t|W^{M},p^M)}{\,d t}
\right|_{t=0}& = -\rE_X D(W_X^{M}\|W_p), \\
\frac{\,d^2 \phi(t|W^{M},p^M)}{\,d t^2}
& \ge 0 .
\end{align*}
Hence, its convexity guarantees
the inequality
$-t\, \rE_X D(W_X^{M}\|W_p) \le\phi(t|W^{M},p^M)$,
which implies the inequality
\begin{align}
\rE_X D(W_X^{M}\|W_p)
\le \frac{\phi(t|W^{M},p^M)}{-t}\Label{1-12-1}
\end{align}
for $0 \,> t \ge -\frac{1}{2}$.

Let $1+s= \frac{1}{1+t}$, then 
$1\ge s \,> 0$ and $t= \frac{-s}{1+s}$.
Since $x \mapsto x^s$ is concave,
\begin{align}
\rE_X (\sum_{j\neq i} W_{X_j}(y))^s 
\le \bigl[\rE_X \sum_{j\neq i} W_{X_j}(y)\bigl]^s 
= (M-1)^s W_p^s(y). \Label{7-4-1}
\end{align}
Using (\ref{7-4-1}) and the relation $(x+y)^s \le x^s + y^s$ for 
two positive real numbers $x,y$, 
we obtain
\begin{align*}
&e^{\phi(t|W^{M},p^M)}
= \sum_y \left(\rE_X (W_X^{M})^{1+s}(y)\right)^{\frac{1}{1+s}}\\
=& \frac{1}{M}\sum_y \Biggl(\rE_X 
\sum_{i=1}^M W_{X_i}(y)
\Bigl(W_{X_i}(y)+ \sum_{j\neq i} W_{X_j}(y)\Bigr)^s
\Biggr)^{\frac{1}{1+s}}\\
\le &\frac{1}{M}\sum_y \Biggl(\rE_X 
\sum_{i=1}^M W_{X_i}(y)
\Bigl(W_{X_i}^s(y)+ \Bigl(\sum_{j\neq i} W_{X_j}(y)\Bigr)^s\Bigr)
\Biggr)^{\frac{1}{1+s}}\\
= &\frac{1}{M}\sum_y 
\Biggl(
\sum_{i=1}^M \rE_X W_{X_i}^{1+s}(y)\\
& \quad 
+\sum_{i=1}^M \rE_X W_{X_i}(y) 
\Bigl(\sum_{j\neq i} W_{X_j}(y)\Bigr)^s
\Biggr)^{\frac{1}{1+s}}\\
\le& \sum_y 
\frac{1}{M}\left(
\sum_{i=1}^M \rE_X W_{X_i}^{1+s}(y)
+\sum_{i=1}^M (M-1)^s W_p^{1+s}(y)
\right)^{\frac{1}{1+s}}\\
=  &\frac{1}{M}\sum_y
\left(M \rE_x W_{x}^{1+s}(y)
+M (M-1)^s W_p^{1+s}(y)
\right)^{\frac{1}{1+s}}\\
\le &
\frac{1}{M}\sum_y 
\left(M \rE_x W_{x}^{1+s}(y)\right)^{\frac{1}{1+s}}
+\left(M (M-1)^s W_p^{1+s}(y)\right)^{\frac{1}{1+s}}\\
=&
\sum_y 
\frac{\left(\rE_x W_{x}^{1+s}(y)\right)^{\frac{1}{1+s}}}{M^{\frac{s}{1+s}}}
+\Bigl(\frac{M-1}{M}\Bigr)^{\frac{s}{1+s}} W_p(y)\\
\le & 1+
\frac{1}{M^{\frac{s}{1+s}}} 
\sum_y \left(\rE_x W_{x}^{1+s}(y)\right)^{\frac{1}{1+s}}
= 1+ M^t e^{\phi(t|W,p)}.
\end{align*}
Since $-t$ is positive, 
the desired inequality (\ref{1-11-3})
follows from (\ref{1-12-1}) and the above inequality.
\endproof

Next, we proceed to the relation with identification codes.
In order to discuss this relation, 
we focus on channel resolvability of the worst input case,
and define the following values:
\begin{align*}
\epsilon(M,W)\defeq&
\max_{p\in \cP(\cX)}
\min_{\Psi:|\Psi|\le M}\epsilon(\Psi,W_p),\\
D(M,W)\defeq&
\max_{p\in \cP(\cX)}
\min_{\Psi:|\Psi|\le M}D(\Psi,W_p),
\end{align*}
which satisfies 
\begin{align}
\epsilon(M,W)\le
2 \max_p \rE_p 
W_x \left\{ \frac{W_x}{W_p}(y) \,> C\right\} + 
\sqrt{\frac{C}{M}} ,
\Label{3-3-3}
\end{align}
for any real number $C\,>0$.
\begin{lem}\Label{H-V}
(Han \& Verd\'{u}\cite{Han-V})
If 
the cardinality $|\cX|$ is finite, and if
an identification code $\Phi$ and 
an integer $M$
satisfy 
\begin{align*}
1- \mu (\Phi)-\lambda(\Phi)
\,> \epsilon (M,W),
\end{align*}
then
\begin{align}
|\cX|^{M} \ge |\Phi|.\Label{3-1}
\end{align}
\end{lem}
\begin{proof}
Let the identification code $\Phi$ be a triplet
$(N, \{Q_1 ,\ldots, Q_N\}, \{\cD_1,\ldots,\cD_N\})$,
then 
there exist $N$ $M$-types $Q_1', \ldots Q_N'$ such that
\begin{align*} 
d(W_{Q_i}, W_{Q_i'}) \le \epsilon (M,W).
\end{align*}
Since 
the inequalities
\begin{align*}
& 2 \epsilon (M,W) + d( W_{Q_i'},  W_{Q_j'} )\\
\ge & d( W_{Q_i},  W_{Q_i'} )+ d(W_{Q_j}, W_{Q_j'} )+ d(W_{Q_i'}, 
W_{Q_j'} ) \\
\ge &
d( W_{Q_i}, W_{Q_j} ) \ge 
2(W_{Q_i}(\cD_i)-  W_{Q_j}(\cD_i)) \\
\ge & 2 (1- \mu (\Phi)-\lambda(\Phi))
\end{align*}
hold for any $i\neq j$,
we can show
\begin{align*}
d(W_{Q_i'}, W_{Q_j'} )\,> 0,
\end{align*}
which implies that $Q_i' $ is different from $Q_j'$.
However, the total number of 
$M$-types
is less than $|\cX|^M$.
Therefore, we obtain (\ref{3-1}).
\end{proof}

\section{Wire-tap channel in non-asymptotic setting}
Next, we discuss the message transmission with the 
wire-tapper who has less information than
the main receiver.
This problem is formulated as follows.
Let $\cY$ be the probability space of the main receiver,
and $\cZ$ be the space of the wire-tapper,
then the main channel from the transmitter to the main receiver
is described by $W^B:x \mapsto W^B_x$,
and the wire-tapper channel from the transmitter to the 
the wire-tapper is described by $W^E:x \mapsto W^E_x$.
In this setting,
the transmitter choose $M$ 
distributions $Q_1, \ldots, Q_M$ on $\cX$,
and he generates $x\in \cX$ subject to $Q_i$
when he wants to send the message $i \in \{1, \ldots, M\}$.
The normal receiver prepares $M$ disjoint subsets
$\cD_1,\ldots, \cD_M$ of $\cY$ and 
judges that a message is $i$ if $y$ belongs to $\cD_i$.
Therefore, the triplet $(M,\{Q_1, \ldots, Q_M\},
\{\cD_1,\ldots, \cD_M\})$ is called a
code, and is described by $\Phi$.
Its performance is given by the following quantities.
One is the size $M$, which is denoted by $|\Phi|$.
The second one is the average 
error probability $\epsilon_B(\Phi)$:
\begin{align*}
\epsilon_B(\Phi)\defeq
\frac{1}{M} \sum_{i=1}^M  W_{Q_i}^B (\cD_i^c),
\end{align*}
and the third one is the wire-tapper's information
regarding the transmitted message $I_E(\Phi)$:
\begin{align*} 
I_E(\Phi) \defeq \sum_i \frac{1}{M} D(  W_{Q_i}^E\| W^E_{\Phi}),\quad
W^E_{\Phi}  \defeq \sum_i \frac{1}{M} W_{Q_i}^E.
\end{align*}
A different measure of the wire-tapper's information is given 
by the average variational 
distance $d_E(\Phi)$:
\begin{align*}
d_E(\Phi)\defeq \frac{1}{M(M-1)} \sum_{i\neq j} d( W_{Q_i}^E,W_{Q_j}^E).
\end{align*}
\begin{thm}\Label{3-6}
There exists a code $\Phi$ for any
integers $L,M$,
any real numbers $C, C' \,> 0$,
and any probability distribution $p$ on $\cX$
such that
\begin{align}
|\Phi| &=M \nonumber \\
\epsilon_B(\Phi) & \le 3
\min_{0\le s\le 1}
(ML)^{s}\sum_y \left( \rE_p (W_x^B(y))^{1/(1+s)}\right)^{1+s} 
\Label{3-8-1}\\
\epsilon_B(\Phi) & \le 3
\left(\rE_p W_x  \left\{ \frac{W_x^B}{W_p^B}(y) \le C'\right\}+
\frac{ML}{C'}\right)\Label{3-8-2}\\
I_E(\Phi) & \le  3
\left(\eta(\delta_{p,W^E,C})+ \delta_{p,W^E,C} \log |\cZ| + 
\frac{\delta_{p,W^E,C}'}{L}
\right)\Label{3-6-2} \\
I_E(\Phi) & \le  3
\min_{0 > t \ge -1/2}
\frac{\log (1+ L^t e^{\phi(t|W^E,p)})}{-t}
\Label{7-1-2} \\
d_E(\Phi) & \le  6
\left(2 \delta_{p,W^E,C} + 
\sqrt{\frac{\delta_{p,W^E,C}'}{L}}\right).\Label{3-8-12} 
\end{align}
\end{thm}

\begin{proof}
We prove Theorem \ref{3-6}
by a random coding method.
Let $X=(X_{l,m})$ be $LM$ independent and identical random
variables subject to the distribution $p$ on $\cX$
for integers $l=1, \ldots, L$ and $m=1,\ldots,M$, and
$\cD_{l,m}'(X)$ be the maximum likelihood decoder
of the code $X_{l,m}$,
then we can evaluate as follows by Gallager upper bound\cite{Gal}.
\begin{align*}
&\rE_X \frac{1}{ML}\sum_{l,m} W^B_{X_{m,l}}(\cD_{l,m}'(X)^c)\\
\le & \min_{0\le s\le 1}
(ML)^{s}\sum_y \left(
\rE_p (W_x^B(y))^{1/(1+s)}\right)^{1+s}.
\end{align*}
Since the maximum likelihood decoder is better than 
the code $\cD_{l,m}''(X)= 
\left\{ \frac{W_x^B}{W_p^B}(y) \,> C'\right\}
\setminus \cup_{(l',m')\neq (l,m)}
\left\{ \frac{W_x^B}{W_p^B}(y) \,> C'\right\}$,
we have another evaluation as
\begin{align*}
&\rE_X \frac{1}{ML}\sum_{l,m} W^B_{X_{m,l}}(\cD_{l,m}'(X)^c)\\
\le 
&\rE_X \frac{1}{ML}\sum_{l,m} W^B_{X_{m,l}}(\cD_{l,m}''(X)^c)\\
\le &
\rE_X 
\frac{1}{ML}\sum_{l,m} W^B_{X_{m,l}}
\left\{ \frac{W_{X_{m,l}}^B}{W_p^B}(y) \le C'\right\} \\
&\quad +
\rE_X 
\frac{1}{ML}\sum_{l,m} W^B_{X_{m,l}}
\sum_{(l',m')\neq (l,m)} 
\left\{ \frac{W_{X_{m',l'}}^B}{W_p^B}(y) \le C'\right\} \\
\le &
\rE_{p,x} 
W^B_{x}
\left\{ \frac{W_x^B}{W_p^B}(y) \le C'\right\} \\
& \quad +
W^B_p
(ML-1) \rE_{p,x} 
\left\{ \frac{W_x^B}{W_p^B}(y) \le C'\right\} \\
\le &
\rE_{p,x} 
W^B_{x}
\left\{ \frac{W_x^B}{W_p^B}(y) \le C'\right\}
+
\frac{ML}{C'}.
\end{align*}
Let $Q_m(X)$ be the uniform distribution on 
$\{X_{1,m},\ldots, X_{L,m}\}$,
$\cD_m(X)$ be $\cup_l \cD_{l,m}'(X)$,
and $\Phi(X)$ be the code
$(M,\{Q_m(X)\},\{\cD_m(X)\})$,
then $\rE_X \epsilon_B(\Phi(X))$ is less than 
the right hand sides of (\ref{3-8-1}) and (\ref{3-8-2})
because the average error probability of $\Phi(X)$
is less than the one of the code 
$(ML,\{X_{m,l}\},\{\cD_{l,m}'(X)\})$.

Since
\begin{align*}
& \sum_{m=1}^M  
\frac{1}{M} D( W_{Q_m(X)}^E\|W^E_{\Phi(X)})
+ D(W^E_{\Phi(X)}\| W^E_p) \\
= & \sum_{m=1}^M  
\frac{1}{M} D( W_{Q_m(X)}^E\|W^E_p),
\end{align*}
we obtain
\begin{align*}
& \rE_X I_E(\Phi(X))
= \rE_X \sum_{m=1}^M  
\frac{1}{M} D( W_{Q_m(X)}^E\|W^E_{\Phi(X)}) \\
\le &
\rE_X \sum_{m=1}^M  
\frac{1}{M} D( W_{Q_m(X)}^E\|W^E_p) \\
\le &
\eta(\delta_{p,W^E,C})+ \delta_{p,W^E,C} \log |\cZ| + 
\frac{\delta_{p,W,C}'}{L},
\end{align*}
where the last inequality follows from
Lemma \ref{lem-1-11}. Similarly,
we can show
\begin{align*}
\rE_X I_E(\Phi(X)) \le\frac{\log (1+ L^t e^{\phi(t|W^E,p)})}{-t}.
\end{align*}
Regarding $d_E(\Phi(X))$,
we can calculate 
\begin{align*}
&\rE_X \frac{1}{M(M-1)} \sum_{i\neq j} d( W^E_{Q_i(X)},W^E_{Q_j (X)}) \\
\le &
\rE_X \frac{1}{M(M-1)} \sum_{i\neq j} d( W^E_{Q_i(X)},W^E_p)
+ d( W^E_{Q_j(X)},W^E_p) \\
=& 2
\rE_X d( W^E_{Q_1(X)},W^E_p) \\
\le & 2 \left(2 \delta_{p,W,C} + \sqrt{\frac{\delta_{p,W,C}'}{L}}\right).
\end{align*}
Using Markov inequality, we obtain
\begin{align*}
\rP_X \{ \epsilon_B(\Phi(X)) \le 3 \rE \epsilon_B(\Phi(X)) \}^c
&\,< \frac{1}{3} \\
\rP_X \{ I_E(\Phi(X)) \le 3 \rE I_E(\Phi(X))\}^c
&\,< \frac{1}{3} \\
\rP_X \{ d_E(\Phi(X)) \le 3 \rE d_E(\Phi(X))\}^c
&\,< \frac{1}{3}.
\end{align*}
Therefore, there exists a code $\Phi$ satisfying desired conditions.
\end{proof} 

\section{General asymptotic setting}
\subsection{Identification code and channel resolvability}
Next, we focus on an arbitrary sequence of channels
$\bW=\{W^n\}_{n=1}^{\infty}$, in which $W^n$ is an arbitrary channel from 
$\cX^n$ to $\cY^n$.
In this setting,
two-types of $(\mu,\lambda)$-identification capacities are defined by
\begin{align}
&D(\mu,\lambda|\bW) \nonumber \\
\defeq &
\sup_{\{\Phi_n\}}
\left\{\left.
\varliminf \frac{1}{n}\log\log |\Phi_n|
\right|
\varlimsup \mu (\Phi_n)\,< \mu,
\varlimsup \lambda(\Phi_n)\le \lambda
\right\}\nonumber\\
& D^{\dagger}(\mu,\lambda|\bW) \nonumber\\
\defeq & 
\sup_{\{\Phi_n\}}
\left\{\left.
\varliminf \frac{1}{n}\log\log |\Phi_n|
\right|
\varliminf \mu (\Phi_n)\,< \mu,
\varlimsup \lambda(\Phi_n)\le \lambda
\right\}.\nonumber
\end{align}
However, in the case of $\mu =0$,
we replace $\varlimsup \mu (\Phi_n)\,< \mu,
(\varliminf \mu (\Phi_n)\,< \mu)$ by $
\varlimsup \mu (\Phi_n)=0,
(\varliminf \mu (\Phi_n)=0)$ at the above two definitions.
On the other hand,
two-types $\epsilon$-resolvability capacities are defined by
\begin{align*}
S(\epsilon|\bW) & \defeq
\sup
\left\{\left.R
\right|
\varlimsup \epsilon(e^{nR},W^n) \le \epsilon
\right\} \\
S^{\dagger}(\epsilon|\bW) & \defeq
\sup
\left\{\left.R
\right|
\varliminf \epsilon(e^{nR},W^n) \le \epsilon
\right\},
\end{align*}
where the case of $\epsilon=2$,
we replace $\le \epsilon $ by $\,< 2$ 
at the above two definitions.

In the information spectrum method,
the following quantities 
are defined for arbitrary sequence $\bp = \{p^n\}_{n=1}^{\infty}$
of input probability distributions:
\begin{align*}
&\overline{I}(\epsilon| \bp, \bW) \\
\defeq &
\inf\left\{
a \left|
\varlimsup \rE_{p^n}
W_x^n\left\{ 
\frac{1}{n}\log \frac{W_x^n}{W_{p^n}^n}(y)\,> a
\right\}
\le \epsilon
\right.
\right\}\\
& \underline{I}(\epsilon| \bp, \bW)\\
\defeq &
\inf\left\{
a \left|
\varliminf \rE_{p^n}
W_x^n\left\{ 
\frac{1}{n}\log \frac{W_x^n}{W_{p^n}^n}(y)\,> a
\right\}
\le \epsilon
\right.
\right\},
\end{align*}
where the case of $\epsilon =1$,
we replace $\le $ by $\,<$ at the above definitions.
These quantities have another expression as
\begin{align*}
&\overline{I}(\epsilon| \bp, \bW) \\
=& \sup \left\{
a \left|
\varliminf 
\rE_{p^n}
W_x^n\left\{ 
\frac{1}{n}\log \frac{W_x^n}{W_{p^n}^n}(y)\le a
\right\}
\,<1- \epsilon
\right.
\right\}, \\
&\underline{I}(\epsilon| \bp, \bW)\\
=& \sup \left\{
a \left|
\varlimsup 
\rE_{p^n}
W_x^n\left\{ 
\frac{1}{n}\log \frac{W_x^n}{W_{p^n}^n}(y)\le a
\right\}
\,<1- \epsilon
\right.
\right\}.
\end{align*}
\begin{thm}\Label{t-3}
Assume that $|\cX^n|= d^n$, then
the above quantities satisfy the following relations.
\begin{align}
& \sup_{\bp} \underline{I}(\epsilon| \bp, \bW) 
\le D (1-\epsilon,0|\bW)
\le S^{\dagger}(\epsilon|\bW) \nonumber \\
\le &\sup_{\bp} \underline{I}(\frac{\epsilon}{2}| \bp, \bW) 
\Label{3-3-1}\\
& \sup_{\bp} \overline{I}(\epsilon| \bp, \bW) 
\le D^{\dagger} (1-\epsilon,0|\bW)
\le S(\epsilon|\bW)\nonumber \\
\le & \sup_{\bp} \overline{I}(\frac{\epsilon}{2}| \bp, \bW) ,
\Label{3-3-2}
\end{align}
for any real number $0 \le \epsilon \,<1$.
However, the first inequalities in (\ref{3-3-1}) and (\ref{3-3-2})
hold for $0 \le \epsilon \le 1$, and 
the third ones hold for $0\le \epsilon \le 2$.
In particular,
we obtain
\begin{align}
\sup_{\bp} \underline{I}(0| \bp, \bW) 
&= D (1,0|\bW)
= S^{\dagger}(0|\bW) \Label{1-13-1}\\
\sup_{\bp} \overline{I}(0| \bp, \bW) 
&= D^{\dagger} (1,0|\bW)
= S(0|\bW),\Label{1-13-2}
\end{align}
which is desired in Han \& Verd\'{u}\cite{Han-V} and 
Han\cite{Han-book}\footnote{Theorem 6 in Han and Verd\'{u}\cite{Han-V}
claims that $S(0|\bW)=\sup_{\bp} \overline{I}(0| \bp, \bW) $
always holds for any channel $\bW$ if the input alphabet is finite.
However, the proof in \cite{Han-V} contains mistake in part, as
is mentioned in section 6.3 in Han\cite{Han-book}.
Therefore, it has been an open problem
as to whether this inequality holds or not.}.
\end{thm}
This theorem indicates the existence of a code satisfying the following:
The second error probability $\lambda$ is asymptotically independent for the 
behavior of the distribution of the random variable of 
likelihood and always goes to $0$, and only the second error 
probability $\mu$ asymptotically depends on it.
\begin{rem}
Steinberg\cite{Steinberg} claims the inequalities 
\begin{align*}
\sup_{\bp} \underline{I}(\epsilon| \bp, \bW) 
&\ge D (\lambda_1,\lambda_2|\bW),\\
\sup_{\bp} \overline{I}(\epsilon| \bp, \bW) 
&\ge D^\dagger (\lambda_1,\lambda_2|\bW)
\end{align*}
for $\lambda_1+\lambda_2 \,< 1- \epsilon$.
If they are proved, 
by combining the above inequalities and Theorem \ref{t-3},
we can prove the equalities of the above inequalities
in the continuous case.
However, it seems that 
his paper has a gap in counting the maximum number of
different pairs of 
a partial response and an $M'$-type measure 
at the proof of Lemma 2,
which is essential for these inequalities.
That is,
he estimated the total number of 
positive functions on $\cX \times \cY$ with the form
\begin{align*}
f(x,y) =\frac{1}{M'}
\sum_{i=1}^{M'}1_{x_i}(x) \sum_{(x',y')\in F}1_{(x',y')} (x,y),
\end{align*}
where $F$ is an arbitrary subset of $\cX \times \cY$.
The total measure of $f$, {\it i.e.},
$\sum_{(x,y)\in \cX \times \cY}f(x,y)$
is not necessarily less than $1$,
while he indicated that it is less than $1$.
Hence, this total number cannot be bounded by 
$|\cX|^{M'}$.
\end{rem}
\begin{proof}
In order to prove the first inequalities,
we choose an arbitrary real number $R \,< \sup_{\bp} 
\underline{I}(\epsilon| \bp, \bW)$ and a sequence of 
input probability distributions $\bp$ such that
$R \,< R'\defeq \underline{I}(1- \mu | \bp, \bW)$.
Substitute $M=e^{nR},C=e^{nR'},\alpha=\beta=1+\frac{2}{n},
\alpha'=\beta'=\frac{1}{n+2},
\tau= \frac{1}{n+2},
\kappa= \frac{\log 2 +1}{\log n}$ in Theorem \ref{t-1},
then the conditions (\ref{3-3-5}) and (\ref{3-3-4}) are satisfied and
$\gamma= \frac{1}{n+2}$.
Thus, there exists an identification code $\Phi_n$ such that
\begin{align*}
|\Phi_n|& = \left\lfloor \frac{e^{\frac{e^{nR}}{n+2}}}{e^{1+nR} }\right\rfloor \\
\mu(\Phi_n)
& \le (1+\frac{2}{n})^2 \rE_{p^n} 
W_x^n\left\{ 
\frac{1}{n}\log \frac{W_x^n}{W_{p^n}^n}(y)\le R'
\right\} \\
\lambda(\Phi_n) 
& \le \frac{\log 2 +1}{\log n}+(n+2)^2 \frac{1}{e^{nR'}} 
\lceil (n+2)e^{nR} \rceil \\
& \cong \frac{\log 2 +1}{\log n}+(n+2)^3 e^{-n(R'-R)} .
\end{align*}
Therefore,
we obtain 
\begin{align}
&\lim \frac{1}{n}\log \log |\Phi_n| = R, \nonumber \\
&\varlimsup \mu(\Phi_n) \le
\varlimsup \rE_{p^n} 
W_x^n\left\{ 
\frac{1}{n}\log \frac{W_x^n}{W_{p^n}^n}(y)\le R'
\right\}\,< \mu \Label{3-2-1}\\
& \lim \lambda(\Phi_n) =0, \nonumber 
\end{align}
which implies that
$D (\mu,0|\bW) \ge R'$.
Thus, we obtain the first inequality in (\ref{3-3-1})
for $0 \le \epsilon \,< 1$.
In the case of $\epsilon = 1$,
we need to replace $\,< \mu$ by $=0$ at (\ref{3-2-1}).
By replacing $\varlimsup$ by $\varliminf$ at (\ref{3-2-1}),
we can similarly prove $D^{\dagger} (\mu,0|\bW) \ge 
\sup_{\bp} \overline{I}(1-\mu| \bp, \bW) $.

Next,
we proceed to the second inequalities.
Let $R$ be an arbitrary real number such that
$R\,> D(1-\epsilon,0|\bW)$.
Then, there exists a sequence $\{\Phi_n\}$
of identification codes
such that
\begin{align*}
R= \varlimsup \frac{1}{n} \log \log |\Phi_n|,
\varlimsup \mu (\Phi_n) < 1-\epsilon,
\lim \lambda (\Phi_n)=0.
\end{align*}
Therefore, we can choose an integer $N$ large enough, 
such that
$1 - \mu(\Phi_n) -\lambda (\Phi_n) \ge
1- \varlimsup \mu (\Phi_n)\,> \epsilon$.
Moreover, we choose a strictly increasing sequence $\{a_n\}$
of integers such that
$a_1\ge N$ and
$1- \varlimsup \mu (\Phi_{a_n})\,> 
\epsilon (e^{a_n R'},W^n)$,
where $R'= S^{\dagger}(\epsilon, \bW)$.

Thus, Lemma \ref{H-V} yields that
$(d^{a_n})^{e^{a_n R'}} \ge |\Phi_{a_n}|$,
which implies that
$R'\ge R$.
We obtain the second inequalities in (\ref{3-3-1}).
We can prove the second inequalities in (\ref{3-3-2})
by choosing a strictly increasing sequence $\{a_n\}$
of integers such that
$1 - \mu(\Phi_{a_n}) -\lambda (\Phi_{a_n}) \ge
1- \varliminf \mu (\Phi_n)\,> \epsilon$.

Finally, we prove the third inequalities
by using another expression of
$\sup_{\bp}\underline{I}(\epsilon| \bp, \bW)
$:
\begin{align*}
& \sup_{\bp}\underline{I}(\epsilon| \bp, \bW) \\
= &\inf \left\{
a \left|
\varlimsup 
\max_{p^n} \rE_{p^n}
W_x^n\left\{ 
\frac{1}{n}\log \frac{W_x^n}{W_{p^n}^n}(y)\,> a
\right\}
\le\epsilon
\right.
\right\}.
\end{align*}
Let $R$ and $R'$ be arbitrary real numbers such that
$R \,>\sup_{\bp}\underline{I}(\epsilon/2| \bp, \bW)$
and 
$R \,>R'\,> \sup_{\bp}\underline{I}(\epsilon| \bp, \bW)$,
then the inequality (\ref{3-3-3}) yields that
\begin{align*}
&\epsilon(e^{n R} ,W^n)\\
\le &
2 \min_{p^n} \rE_{p ^n}
W_x^n \left\{ \frac{1}{n} \log \frac{W_x^n}{W_{p^n}^n}(y) \,>
R'\right\} + e^{-n (R-R')/2}.
\end{align*}
Taking the limit $\varliminf$,
we obtain
\begin{align}
\varliminf \epsilon(e^{n R} ,W^n)
\le \epsilon, \Label{3-3-6}
\end{align}
which implies 
$S^{\dagger}(2 \epsilon, \bW) \le R$.
Thus, we obtain the third inequality in (\ref{3-3-1})
for $0 \le \epsilon \,< 1$.
In the case of $\epsilon =2$,
we need to replace $\le \epsilon$ by $\,< 1$
at (\ref{3-3-6}).
By replacing $\varliminf$ by $\varlimsup$ in the above,
we can prove the third one in (\ref{3-3-2}).
\end{proof}

\subsection{Wire-tap channel}
Next, we focus on a general sequence 
$(\bW^B=\{W^{B,n}\}, \bW^E=\{W^{E,n}\})$  
of 
wire-tap channels, and define the 
following two kinds of capacities
by
\begin{align*}
&C_d(\bW^B,\bW^E ) \\
\defeq &
\sup_{\{\Phi_n\}}
\left\{\left.
\varliminf \frac{1}{n}\log |\Phi_n|
\right|
\lim \epsilon_B (\Phi_n)=
\lim d_E(\Phi_n)= 0
\right\} \\
&C_I (\bW^B,\bW^E ) \\
\defeq &
\sup_{\{\Phi_n\}}
\left\{\left.
\varliminf \frac{1}{n}\log |\Phi_n|
\right|
\lim \epsilon_B (\Phi_n)=
\lim \frac{I_E(\Phi_n)}{n}= 0
\right\} .
\end{align*}

\begin{lem}\Label{t-1-11}
The inequality
\begin{align}
C_d(\bW^B,\bW^E) 
\ge \underline{I}(1|\bp,\bW^B)
- \overline{I}(0|\bp,\bW^E)
\Label{3-8-13}
\end{align}
holds for any sequence of input distributions $\bp=\{p^n\}$.
Furthermore, if $|\cZ^n|= d^n$,
\begin{align}
C_I(\bW^B,\bW^E) 
\ge \underline{I}(1|\bp,\bW^B)
- \overline{I}(0|\bp,\bW^E) \Label{3-8-13-1}.
\end{align}
\end{lem}
This theorem is an information spectrum version of Wyner's result
\cite{Wyner},
that will be mentioned in the next section.

\begin{proof}
Let $R' \,> \overline{I}(0|\bp,\bW^E)$,
$R\,< \underline{I}(1|\bp,\bW^B)- R'$ and 
choose a real number $a$ such that $0\,< a \,< 
\min \{\underline{I}(1|\bp,\bW^B) -(R+R'), 
R'- \overline{I}(0|\bp,\bW^E)\}$.
Substituting $M= e^{n R}, L=e^{n R'},
C= e^{n(R'-a)}, C'=e^{n(R+R'+a)}$,
we can show that the right hand side of (\ref{3-8-2}) goes to $0$,
and that 
\begin{align*}
\delta_{p^n,W^{E,n},e^{n(R'-a)}}  \to 0 , \quad
\frac{\delta_{p^n,W^{E,n},e^{n(R'-a)}}'}{e^{n R'}}  \to 0.
\end{align*}
Hence, 
the right hand side of (\ref{3-8-12}) go to $0$.
Concerning (\ref{3-6-2}),
the relations
\begin{align*}
& \frac{1}{n}
\Bigl( 
\eta(\delta_{p^n,W^{E,n},e^{n(R'-a)}})+ \delta_{p,W^{E,n},e^{n(R'-a)}} 
\log |\cZ^n| \\
& \quad + \frac{\delta_{p^n,W^{E,n},e^{n(R'-a)}}'}{e^{n R'}}
\Bigr)\\
= &
\frac{1}{n}
\eta(\delta_{p^n,W^{E,n},e^{n(R'-a)}})+ \delta_{p,W^E,e^{n(R'-a)}}
\log d \\
& \quad + 
\frac{1}{n}
\frac{\delta_{p^n,W^{E,n},e^{n(R'-a)}}'}{e^{n R'}}\\
\to & 0
\end{align*}
hold. Therefore, we obtain (\ref{3-8-13}) and (\ref{3-8-13-1}).
\end{proof}

Conversely, we obtain the following lemma.
\begin{lem}\Label{l-1-14}
Let $\bQ= \{Q^n\}$ be a sequence of channels from 
arbitrary set $\tilde{\cX}^n$ to the set
$\cX^n$ and $\bp=\{p^n\}$ be a sequence of
distributions on $\tilde{\cX}^n$.
Then, the inequalities
\begin{align}
C_d(\bW^B,\bW^E) 
\le &
\sup_{\bp, \bQ}
\left\{
\underline{I}(1|\bp,\bW^B \bQ)- \overline{I}(0|\bp,\bW^E\bQ)
\right\}\Label{1-14-4}\\
C_I(\bW^B,\bW^E)
\le &
\sup_{\bp, \bQ}
\left\{
\underline{I}(1|\bp,\bW^B \bQ)- \overline{I}(0|\bp,\bW^E\bQ)
\right\}\Label{1-14-5}
\end{align}
hold, where
$\bW \bQ =\{W^{n} Q^n \}$ denotes the sequence of channels 
from $\tilde{\cX}^n$ to $\cY^n$:
\begin{align*}
(W^n Q^n)_{\tilde{x}}(y)\defeq \sum_{x \in \cX^n} W^n_{x}(y) 
Q^n_{\tilde{x}}(x)
\end{align*}
for a sequence of channels $\bW=\{W^n\}$ 
from $\cX^n$ to $\cY^n$.
\end{lem}

Hence, applying Lemma \ref{t-1-11} to the sequence of the channels 
$\bW^B \bQ, \bW^E\bQ$,
we obtain the following theorem.
\begin{thm}\Label{t-1-17}
\begin{align*}
&C_d(\bW^B,\bW^E) 
=C_I(\bW^B,\bW^E)  \\
=& \sup_{\bp, \bQ}
\left\{
\underline{I}(1|\bp,\bW^B \bQ)- \overline{I}(0|\bp,\bW^E\bQ)\right\}.
\end{align*}
\end{thm}

\quad
{\it Proof of Lemma \ref{l-1-14}:}
Let $\{\Phi_n =(M_n,\{Q_1^n, \ldots, Q_{M_n}^n\},
\{\cD_1^n, \ldots, \cD_{M_n}^n\})\}$ 
be a sequence of codes of wire-tap channel such that
\begin{align*}
R= \varliminf \frac{1}{n}\log |\Phi_n|,
\quad
\lim \epsilon_B (\Phi_n) =0, \quad
\lim d_E (\Phi_n) =0.
\end{align*}
Hence, Verd\'{u}-Han's result \cite{V-H} yields that
the transmission capacity of the sequence of channel 
$\bW^B \bQ$ is less than $\underline{I}(1|\bp,\bW^B \bQ)$,
which implies 
\begin{align*}
R \le \underline{I}(1|\bp,\bW^B \bQ) .
\end{align*}
Furthermore,
the property $\lim d_E (\Phi_n) =0$ implies that
$S(0|\bW^E\bQ)= 0$.
Hence, we have
\begin{align}
\overline{I}(0|\bp,\bW^E\bQ)= 0. \Label{1-17-4}
\end{align}
Thus, we obtain
\begin{align*}
R = \underline{I}(1|\bp,\bW^B \bQ)- \overline{I}(0|\bp,\bW^E\bQ),
\end{align*}
which implies (\ref{1-14-4}).

Next, we assume that
a sequence of codes of wire-tap channel 
$\{\Phi_n =(M_n,\{Q_1^n, \ldots, Q_{M_n}^n\},
\{\cD_1^n, \ldots, \cD_{M_n}^n\})\}$ 
satisfies that
\begin{align*}
R= \varliminf \frac{1}{n}\log |\Phi_n|,
\quad
\lim \epsilon_B (\Phi_n) =0, \quad
\lim \frac{I_E (\Phi_n)}{n} =0.
\end{align*}
Since the mutual information 
\begin{align*}
I_E (\Phi_n)= \sum_{i=1}^{M_n} 
\frac{1}{M_n} \rE_{(W^{E,n}Q^n)_{i},y}
\log \frac{(W^{E,n}Q^n)_{i}}{
\sum_{i=1}^{M_n} \frac{1}{M_n}(W^{E,n}Q^n)_{i}}(y)
\end{align*}
can be regarded as KL-divergence,
Lemma \ref{l-1-17} yields that
\begin{align*}
& \sum_{i=1}^{M_n} 
\frac{1}{M_n}
(W^{E,n}Q^n)_{i}
\left\{
\frac{1}{n}\log \frac{(W^{E,n}Q^n)_{i}}{
\sum_{i=1}^{M_n} \frac{1}{M_n}(W^{E,n}Q^n)_{i}}(y)
\ge a \right\} \\
& \le 
\frac{I_E (\Phi_n) + \frac{1}{e}}{n a} \to 0
\end{align*}
for any $a\,> 0$.
Thus, we obtain (\ref{1-17-4}).
Therefore, similarly to (\ref{1-14-4}), we obtain (\ref{1-14-5}).
\endproof

\begin{lem}\Label{l-1-17}
Assume that $p$ and $q$ are two probability distributions
on $\Omega$.
Then, we have
\begin{align}
D(p\|q)+ \frac{1}{e}
\ge \alpha 
\cdot p \left\{ \log \frac{p(\omega)}{q(\omega)} \ge \alpha \right\}
.\Label{1-17-1}
\end{align}
\end{lem}
\begin{proof}
We focus on the two probability distributions on 
$\Omega_0 \defeq\left\{ \log \frac{p}{q}(\omega) \,< \alpha \right\}$:
\begin{align*}
p_0(\omega) \defeq \frac{p(\omega)}{p\{\Omega_0\}} , \quad
q_0(\omega) \defeq \frac{q(\omega)}{q\{\Omega_0\}}.
\end{align*}
Hence,
\begin{align*}
&D(p\|q) = 
\sum_{\omega \in \Omega_0^c} p(\omega) \log \frac{p(\omega)}{q(\omega)}
+ \sum_{\omega \in \Omega} p(\omega) \log \frac{p(\omega)}{q(\omega)} \\
\ge & \alpha p\{\Omega_0^c\}
+ \sum_{\omega \in \Omega} p_0(\omega) 
\left(\log \frac{p\{\Omega_0\}}{q\{\Omega_0\}}
+ \log \frac{p_0(\omega)}{q_0(\omega)}\right) \\
= & \alpha p\{\Omega_0^c\}
+ p\{\Omega_0\}\log \frac{p\{\Omega_0\}}{q\{\Omega_0\}}
+ D(p_0\|q_0) \\
\ge & \alpha p\{\Omega_0^c\}
+ p\{\Omega_0\}\log \frac{p\{\Omega_0\}}{q\{\Omega_0\}}
\ge \alpha p\{\Omega_0^c\}
+ p\{\Omega_0\}\log p\{\Omega_0\}.
\end{align*}
Finally, the convexity of the map $x \mapsto x \log x $ guarantees that
$p\{\Omega_0\}\log p\{\Omega_0\}\ge -\frac{1}{e}$.
We obtain (\ref{1-17-1}).
\end{proof}

\section{Exponents in stationary memoryless channel}
\subsection{Channel resolvability}
Next, we proceed to 
the stationary memoryless 
channel of a given channel $W$ as a special case.

First, we treat channel resolvability.
As was shown by 
Han \& Verd\'{u}\cite{Han-V} and Han \cite{Han-book},
the information spectrum quantities of discrete memoryless channel of $W$
is calculated as
\begin{align*}
\sup_{\bp}\overline{I}(\epsilon|\bp,\bW)
=\sup_{\bp}\underline{I}(\epsilon|\bp,\bW)
= \max_p I(p;W)
\end{align*}
for $1\ge \epsilon \ge 0$,
where
\begin{align*}
I(p;W) & \defeq \rE_p D(W_x\|W_p).
\end{align*}
Hence, Theorem \ref{t-3} yields 
\begin{align*}
S(\epsilon|\bW)
=S^{\dagger}(\epsilon|\bW)
= \max_p I(p;W),
\end{align*}
which has been obtained by Han \& Verd\'{u}\cite{Han-V}.
Furthermore, using Theorem \ref{t-2},
we can discuss these problems 
in more details by treating the following 
optimal exponents:
\begin{align*}
& e_\epsilon (R|W,p) \\
\defeq &
\sup_{\{\Psi_n\}} \left\{\left. 
\varliminf \frac{-1}{n}\log \epsilon (\Psi_n,W^n_{p^n}) 
\right|
\varlimsup \frac{1}{n}\log |\Psi_n|\le R
\right\} \\
& e_D (R|W,p) \\
\defeq &
\sup_{\{\Psi_n\}} \left\{\left. 
\varliminf \frac{-1}{n}\log D (\Psi_n, W^n_{p^n}) 
\right|
\varlimsup \frac{1}{n}\log |\Psi_n|\le R
\right\} ,\\
\end{align*}
and 
\begin{align*}
e_\epsilon (R|W) &\defeq \varliminf \frac{-1}{n}\log \epsilon (e^{nR},W^n) ,\\
e_D (R|W) &\defeq \varliminf \frac{-1}{n}\log D (e^{nR},W^n) ,
\end{align*}
where $p^n$ is the $n$-fold identical independent distribution of
$p$.
As is discussed by Oohama \cite{Oohama},
by using Lemma \ref{H-V}, the exponent $e_{\epsilon}(R,W)$
gives a lower bound of strong converse exponent of 
identification code.
\begin{thm}\Label{t-5}
Assume that the cardinality $|\cY|$ is finite, then
\begin{align}
e_\epsilon (R|W,p) & \ge
\max_{1\ge s \ge 0} \left\{\frac{-\psi(s|W,p)+ sR}{1+s} \right\}
\Label{3-8-7}\\
e_D (R|W,p) & \ge
\max_{0 \ge t \ge -1/2}\left\{ - \phi(t|W,p) - tR\right\}
\Label{3-8-8} \\
e_\epsilon (R|W) & \ge
\max_{1\ge s \ge 0} \left\{ \frac{-\psi(s|W)+ sR}{1+s} \right\}\Label{3-9-2}\\
e_D(R|W) & \ge
\max_{0 \ge t \ge -1/2} \left\{ - \max_p \phi(t|W,p) - tR \right\},
\Label{7-1-1}
\end{align}
where $\psi(s|W,p)\defeq \log \rE_p \sum_y W_x^{1+s}(y)W_p^{-s}(y)$
and $\psi(s|W)\defeq \log \max_p \sum_y 
\left(\rE_p W_x^{1+s}(y)\right)^{1-s}$.
\end{thm}
Using Pinsker's inequality
$D(p\|q)\ge \|p-q\|^2$,
we obtain two inequalities
$\frac{1}{2}e_D (R|W,p) \le e_\epsilon (R|W,p)$ and 
$\frac{1}{2}e_D(R|W) \le e_\epsilon (R|W)$,
which implies different lower bounds of exponents:
\begin{align}
e_\epsilon (R|W,p) & \ge
\frac{1}{2} \max_{0 \ge t \ge -1/2}  \left\{- \phi(t|W,p) - tR\right\}
\Label{7-1-5}\\
e_\epsilon (R|W) & \ge
\frac{1}{2}
\max_{0 \ge t \ge -1/2}  \left\{- \max_p \phi(t|W,p) - tR \right\}
.\Label{7-1-6}
\end{align}
We can derive different lower bounds of $e_D (R|W,p)$ and $e_D (R|W)$
from the inequality (\ref{7-1-3}).
However, these bounds are smaller than the bound presented here.

\begin{rem}
Arimoto's strong converse exponent \cite{Arimoto} of 
channel coding of transmission code equals
\begin{align*}
\max_{0 \ge t \ge -1}  \left\{- \max_p \phi(t|W,p) - tR\right\},
\end{align*}
which is a bit greater than the RHS of (\ref{3-8-8})
when $R$ is sufficiently large.

\end{rem}
\begin{rem}
By using inequality (\ref{3-3-10}) and type method,
Oohama \cite{Oohama} has obtained a lower bound of $e_\epsilon (R|W)$:
\begin{align*}
\frac{1}{2}
\max_{0 \ge t \ge -1}  \left\{- \max_p \phi(t|W,p) - tR\right\},
\end{align*}
which is a bit better than (\ref{7-1-6})
when $R$ is sufficiently large.
It is interesting that his approach
is in contrast to our approach to (\ref{7-1-6}),
which is based on (\ref{6-26-1}) not on (\ref{3-3-10}).
\end{rem}
\begin{rem}
It is difficult to treat the exponent of 
the sum of two error probabilities in identification code
based on Theorem \ref{t-1}.
For this purpose, we need a modified version of Theorem \ref{t-1}.
\end{rem}

The following lemma is a preparation of our proof of Theorem \ref{t-5}.

\begin{lem}\Label{3-9-1}
For any $s \ge 0$ and $0 \ge t \,> -1$,
the equalities
\begin{align}
& \max_{p \in \cP(\cX^n)} 
\sum_{y^n \in \cY^n} \left (\rE_{p} (W_x^n(y^n))^{1+s} \right)^{1-s}
\nonumber\\
= &
\left(\max_{p\in \cP(\cX)} 
\sum_y \left( \rE_p W_x^{1+s}(y) \right)^{1-s}
\right)^n \label{6-27-1}\\
& \max_{p \in \cP(\cX^n)} 
\sum_{y^n \in \cY^n} \left (\rE_{p} (W_x^n(y^n))^{\frac{1}{1+t}} \right)^{1+t}
\nonumber\\
= &
\left(\max_{p\in \cP(\cX)} 
\sum_y \left( \rE_p W_x^{\frac{1}{1+t}}(y) \right)^{1+t}
\right)^n \label{7-1-4}
\end{align}
hold.
\end{lem}
\begin{proof}
Since (\ref{7-1-4}) has been shown by Arimoto \cite{Arimoto}, 
we prove only (\ref{6-27-1})
by the same method.
Since the function $f: p \mapsto 
\max_{p\in \cP(\cX)} 
\sum_y \left( \rE_p W_x^{1+s}(y) \right)^{1-s}$ is continuous and
convex function,
if and only if
$f(p^*)= \max_p f(p)$,
there exists a constant $\lambda$ such that
\begin{align*}
&\sum_y W_x^{1+s}(y) \left(
\sum_x p^*(x) W_x^{1+s}(y)\right)^{-s}\\
=& \frac{\partial f}{\partial p(x)}
\left\{
\begin{array}{cc}
= \lambda & \hbox{if } p^*(x)\,> 0 \\
\le \lambda & \hbox{if } p^*(x)= 0 
\end{array}
\right.
\end{align*}
Indeed, $\lambda$ is calculated as
\begin{align*}
\sum_x p(x) \lambda 
&=
\sum_x p(x) 
\sum_y W_x^{1+s}(y) \left(
\sum_x p^*(x) W_x^{1+s}(y)\right)^{-s}\\
&=
\left(
\sum_x p^*(x) W_x^{1+s}(y)\right)^{1-s}.
\end{align*}
Thus,
if and only if
$f(p^*)= \max_p f(p)$,
\begin{align}
&\sum_y W_x^{1+s}(y) \left(
\sum_x p^*(x) W_x^{1+s}(y)\right)^{-s}\nonumber \\
&\left\{
\begin{array}{cc}
= \left(
\sum_x p^*(x) W_x^{1+s}(y)\right)^{1-s} & \hbox{if } p^*(x)\,> 0 \\
\le \left(
\sum_x p^*(x) W_x^{1+s}(y)\right)^{1-s}& \hbox{if } p^*(x)= 0 
\end{array}
\right.\nonumber,
\end{align}
$p^*$ gives the maximum.
Hence, if $p^*$ satisfies the above condition,
$(p^*)^n$ also satisfies the following condition:
\begin{align*}
&\sum_{y^n} (W_{x^n}^n)^{1+s}(y^n) \left(
\sum_{x^n} (p^*)^n(x^n) (W_{x^n}^n)^{1+s}(y^n)\right)^{-s}\nonumber \\
&\left\{
\begin{array}{cc}
= \left(
\sum_{x^n} (p^*)^n(x^n) (W_{x^n}^n)^{1+s}(y^n)\right)^{1-s}
& \hbox{if } (p^*)^n(x^n)\,> 0 \\
\le \left(
\sum_{x^n} (p^*)^n(x^n) (W_{x^n}^n)^{1+s}(y^n)\right)^{1-s}
& \hbox{if } (p^*)^n(x^n)= 0 ,
\end{array}
\right. 
\end{align*}
which is a necessary and sufficient condition for 
\begin{align*}
& \sum_{y^n \in \cY^n} 
\left(\rE_{(p^*)^n} (W_x^n(y^n))^{1+s} \right)^{1-s}\\
= &
\max_{p \in \cP(\cX^n)} 
\sum_{y^n \in \cY^n} 
\left(\rE_{p} (W_x^n(y^n))^{1+s} \right)^{1-s}.
\end{align*}
It implies the equation (\ref{6-27-1}).
\end{proof}

\quad{\it Proof of Theorem \ref{t-5}:}
By inequality (\ref{6-26-1}) of Theorem \ref{t-2},
we have
\begin{align}
D(e^{nR},W^^n_{p^n})
&\le
\frac{\log (1+ (e^{nR})^t e^{ \phi(t|W^n,p^n)})}{-t}\nonumber \\
& \le
\frac{(e^{nR})^t e^{ \phi(t|W^n,p^n)}}{-t}\nonumber \\
& =
\frac{e^{n( \phi(t|W,p)+t R)}}{-t},\label{A1}
\end{align}
for $0 > t \ge -1/2$, where 
the second inequality follows from $\log (1+x) \le x$.
From (\ref{A1}), we obtain 
\begin{align}
e_D(R|W,p)\ge -\phi(t|W,p)-t R
\end{align}
for $0 > t \ge -1/2$.
Since $\phi(t|W,p)+ tR$ is 
continuous for $t$,
the inequality (\ref{3-8-8}) holds.
By inequality (\ref{6-26-1}) and Lemma \ref{3-9-1},
we have
\begin{align}
D(e^{nR},W^n)
& \le
\max_{p \in \cP(\cX^n)}
\frac{\log (1+ (e^{nR})^t e^{ \phi(t|W^n,p)})}{-t}\nonumber \\
& \le
\frac{
(e^{nR})^t \cdot
\max_{p \in \cP(\cX^n)}
e^{ \phi(t|W^n,p)}
}{-t} \nonumber \\
= &
\frac{
(e^{nR})^t \cdot
\max_{p \in \cP(\cX)}
e^{n \phi(t|W,p)}
}{-t} \label{A2}
\end{align}
for $0 > t \ge - 1/2$.
Hence, in a manner similar to the derivation of 
(\ref{3-8-8}) from (\ref{A1}), we obtain (\ref{7-1-1}) from (\ref{A2}).

Next, we derive (\ref{3-8-7}) and (\ref{3-9-2}).
To this end, we first derive an upper bound of 
\begin{align*}
2\delta_{p,W,e^{R'}}+
\sqrt{
\frac{\delta'_{p,W,e^{R'}}}
{e^R}}.
\end{align*}
For any $1\ge s \ge 0$,
we choose $R'\defeq \frac{\psi(s|W,p)+R}{1+s}$.
By using Markov inequality, we can evaluate 
$\delta_{p,W,e^{R'}}$ and $\delta_{p,W,e^{R'}}'$ as
\begin{align}
& \delta_{p,W,e^{R'}}
\le \rE_p \sum_{y \in 
\left\{ \frac{W_x}{W_p}(y) \,> e^{R'} \right\}}
W_x (y) \left(\frac{e^{- R'} W_x(y)}{W_p(y)}\right)^s
\nonumber\\
\le &
\rE_p \sum_{y} 
W_x (y) \left(\frac{ W_x(y)}{W_p(y)}\right)^s e^{- s R'} 
=e^{ \psi(s|W,p)-s R'}\nonumber\\
=& e^{\frac{ \psi(s|W,p)-sR}{1+s}}\label{A3}
\end{align}
and
\begin{align}
& \delta_{p,W,e^{R'}}'
\le \rE_p \sum_{y \in 
\left\{ \frac{W_x}{W_p}(y) \le e^{R'} \right\}}
\frac{W_x (y)^2}{W_p (y)} 
\left(\frac{W_p(y)}{e^{- R'} W_x(y)}\right)^{1-s}\nonumber\\
\le &
\rE_p \sum_{y} 
W_x (y) \left(\frac{ W_x(y)}{W_p(y)}\right)^s e^{(1- s) R'} 
=e^{ \psi(s|W,p)+(1- s) R'},\label{A4}
\end{align}
respectively.
Inequality (\ref{A4}) yields
\begin{align}
\sqrt{\frac{\delta_{p,W,e^{R'}}'}{e^{R}}}
\le e^{ \frac{\psi(s|W,p)+(1- s) R'-R }{2}}= 
e^{\frac{ \psi(s|W,p)-sR}{1+s}}.\label{A5}
\end{align}
Combining (\ref{A3}) and (\ref{A5}), 
we have
\begin{align}
2\delta_{p,W,e^{R'}}+
\sqrt{
\frac{\delta'_{p,W,e^{R'}}}
{e^R}}
\le 3 e^{\frac{ \psi(s|W,p)-sR}{1+s}}.\label{A6}
\end{align}
Hence, (\ref{A6}) and (\ref{3-3-10}) in Theorem \ref{t-2}
guarantee that
\begin{align}
\epsilon(e^{nR},W^n_{p^n})
\le 3 e^{\frac{ \psi(s|W^n,p^n)-snR}{1+s}}
= 3 e^{n\frac{ \psi(s|W,p)-sR}{1+s}}\label{A7}
\end{align}
for $1 \ge s \ge 0$
because $\psi(s|W^n,p^n)= n \psi(s|W,p)$.
Thus, (\ref{A7}) implies that
\begin{align}
e_\epsilon(R|W,p)
\ge 
\frac{-\psi(s|W,p)+sR}{1+s}\label{A8}
\end{align}
for $1 \ge s \ge 0$.
Taking the maximum for $1 \ge s \ge 0$, 
we obtain (\ref{3-8-7}).


We proceed to the proof of (\ref{3-9-2}).
By inequalities (\ref{3-3-10}) and (\ref{A6}), we obtain 
\begin{align}
\epsilon(e^{nR},W^n)\le
\max_{p \in \cP(\cX^n)}
3 e^{\frac{\psi(s|W^n,p)-snR}{1+s}}\nonumber \\
=
3 e^{\frac{-snR}{1+s}}
\left[
\max_{p \in \cP(\cX^n)}
e^{\frac{\psi(s|W^n,p)}{1+s}}
\right]^{\frac{1}{1+s}}\label{A9}.
\end{align}
Now, we estimate an upper bound of 
\begin{align}
e^{\psi(s|W^n,p)}
= \rE_p\sum_y W_x^{1+s}(y)W_p^{-s}(y).
\end{align}
Since the map $x \mapsto x^{1+s}$ is convex,
we have
\begin{align*}
W_p(y) = \rE_p W_x(y) \ge
\rE_p W_x^{1+s}(y),
\end{align*}
which imply that
\begin{align*}
W_p^{-s} (y) \le \left(\rE_p (W_x(y))^{1+s}\right)^{-s}.
\end{align*}
Hence, the relations
\begin{align}
& \rE_p \sum_y W_x^{1+s}(y)W_p^{-s} (y)
= \sum_y \rE_p W_x^{1+s}(y)W_p^{-s} (y)\nonumber \\
\le & \sum_y \left (\rE_p W_x^{1+s}(y) \right)^{1-s}\Label{3-12-1}
\end{align}
hold. Using (\ref{3-12-1}) and Lemma \ref{3-9-1}, 
we can evaluate
\begin{align}
&\max_{p \in \cP(\cX^n)} 
\rE_{p} \sum_{y\in \cY^n} W_x^n(y)^{1+s}
W_{p}^n (y)^{-s} \nonumber \\
\le &
\max_{p \in \cP(\cX^n)} 
\sum_{y \in \cY^n} \left (\rE_{p} (W_x^n(y))^{1+s} \right)^{1-s}
\nonumber \\
= &
\left(\max_{p\in \cP(\cX)} 
\sum_y \left( \rE_p W_x^{1+s}(y) \right)^{1-s}
\right)^n = e^{n (\psi(s|W))}.
\label{A10}
\end{align}
Combining (\ref{A9}) and (\ref{A10}), 
we have
\begin{align}
\epsilon(e^{nR}, W^n)
\le e^{n \frac{\psi(s|W)-sR}{1+s}},
\label{A11}
\end{align}
for any $1 \ge s \ge 0$.
In a manner similar to the derivation of (\ref{3-8-7})
from (\ref{A7}), we can derive (\ref{3-9-2}) from (\ref{A11}).
\endproof

\subsection{Wire-tap channel}
Next, we proceed to discrete memoryless wire-tap channel.
Applying Theorem \ref{t-1-11} 
to this case with the input identical and independent distribution,
we obtain 
\begin{align*}
& C(\bW^B,\bW^E) \\
\defeq &
\sup_{\{\Phi_n\}}
\left\{\left.
\varliminf \frac{1}{n}\log\log |\Phi_n|
\right|
\lim \epsilon_B (\Phi_n)=
\lim d_E(\Phi_n)= 0
\right\} \\
\ge &\sup_p \left\{I(p;W^B)- I(p;W^E)\right\},
\end{align*}
which has been obtained by Wyner \cite{Wyner}.
Hence, Theorem \ref{t-1-11} can be 
regarded as a general extension of Wyner's result.
Moreover, using Lemma \ref{3-6},
we derived several explicit lower bounds of exponents.
\begin{thm}\Label{t-7}
Assume that the cardinality $|\cZ|$ is finite,
then there exists a sequence $\{\Phi_n\}$ of codes
for any real numbers $R, R'$ and any probability distribution $p$
such that
\begin{align}
\lim \frac{1}{n}\log |\Phi_n|& = R \nonumber \\
\varliminf \frac{-1}{n}\log \epsilon_B(\Phi_n) & \ge
\max_{1 \ge s \ge 0}  \left\{-\phi(s|W^B,p) - s(R+R')\right\}\Label{3-8-10} \\
\varliminf \frac{-1}{n}\log I_E(\Phi_n) & \ge
\max_{0\ge t \ge -1/2} \left\{ -\phi(t|W^E,p)- tR'\right\} \Label{6-8-10} \\
\varliminf \frac{-1}{n}\log d_E(\Phi_n) & \ge
\max_{1\ge s \ge 0} \left\{\frac{-\psi(s|W^E,p)+ sR'}{1+s}\right\}
\Label{7-8-10}\\
\varliminf \frac{-1}{n}\log d_E(\Phi_n) &\ge
\frac{1}{2} \max_{0\ge t \ge -1/2}  \left\{-\phi(t|W^E,p)- tR'\right\}.
\Label{1-12-3}
\end{align}
Indeed, these exponents are very useful for evaluating
error and wire-tapper's information
for a finite $n$.
\end{thm}
\begin{proof}
The inequality (\ref{3-8-10}) immediately follows from
(\ref{3-8-1}).
By using an evaluation similar to (\ref{3-8-8}),
we can show (\ref{6-8-10}) from (\ref{3-6-2}).
Furthermore, by using an evaluation similar to (\ref{3-8-7}),
we can show (\ref{7-8-10}) from (\ref{3-8-12}).
\end{proof}
\section{Comparison of lower bounds of exponents}
Finally, we compare the lower bounds
(\ref{3-8-7}), (\ref{3-9-2}), (\ref{7-1-5}), and (\ref{7-1-6})
of error exponents of channel resolvability.

\begin{thm}
Assume that $\Delta \defeq R- I(p;W)$ is sufficiently small.
Then, 
RHSs of (\ref{3-8-7}) and (\ref{7-1-5}) 
(which are lower bounds of exponent of the variational distance)
are approximately
calculated as
\begin{align*}
\hbox{RHS of } (\ref{3-8-7})~&
\max_{1\ge s \ge 0} \left\{ \frac{-\psi(s|W,p)+ sR}{1+s} \right\}
 \cong \frac{\Delta^2}{4 J(p;W)}\\
\hbox{RHS of } (\ref{7-1-5})~&
\frac{1}{2}\max_{0 \ge t \ge -1/2} \left\{ - \phi(t|W,p) - tR\right\} 
 \cong \frac{\Delta^2}{8 J(p;W)}, 
\end{align*}
where 
\begin{align*}
&J(p;W) \\
\defeq 
&\frac{1}{2} 
\left( 
\rE_{p,x} \rE_{W_x,y}
(\log W_x(y) - \log W_p(y))^2- I^2(p;W) \right) .
\end{align*}
Moreover, 
RHSs of (\ref{3-9-2}) and (\ref{7-1-6}) 
(which are lower bounds of exponent of the worst variational distance)
are approximately
calculated as
\begin{align*}
\hbox{RHS of } (\ref{3-9-2})~
&\max_{s \ge 0}\left\{
\frac{-\psi(s|W)+sR}{1+s}\right\}\\
& \cong \frac{\Delta^2}{4 (J(p_0;W)+\rE_{p_0} H(W_x))}, \\
\hbox{RHS of } (\ref{7-1-6})~
&\frac{1}{2}\max_{0 \ge t \ge -1/2} \left\{- \max_p\phi(t|W,p) - tR\right\}\\
& \cong \frac{\Delta^2}{8 J(p_0;W) },
\end{align*}
where $p_0 \defeq \argmax_p I(p;W)$.
\end{thm}
Thus, when $R$ is sufficiently close to
$\max_p I(p;W)$,
(\ref{3-8-7}) gives a better lower bound than
(\ref{7-1-5}).
Of course, this comparison can be applied to 
exponents of eavesdropper's information in wire-tap channel,
{\it i.e.},
the comparison of RHSs of (\ref{7-8-10}) and (\ref{1-12-3}).
On the other hand, 
(\ref{3-9-2}) gives a better lower bound than
(\ref{7-1-6}),
if and only if
\begin{align*}
&\rE_{p_0} H(W_x)\\
&\le \frac{1}{2} 
\left( 
\rE_{p,x} \rE_{W_x,y}
(\log W_x(y) - \log W_p(y))^2
- I^2(p;W) \right).
\end{align*}
Therefore, although $R- \max_p I(p;W)$ is small enough,
the relation between 
bounds (\ref{3-9-2}) and (\ref{7-1-6}) is not clear.

\begin{proof}
By using a Taylor expansion, we obtain the approximations:
\begin{align*}
\psi(s|W,p)&\cong  I(p;W)s + J(p;W)s^2\\
\phi(t|W,p)&\cong - I(p;W)t + J(p;W)t^2\\
\psi(s|W) &\cong I(p_0;W) s + (J(p_0;W)+\rE_{p_0} H(W_x)) s^2,
%
\end{align*}
Thus,
\begin{align*}
&\max_{1\ge s \ge 0} \left\{ \frac{-\psi(s|W,p)+ sR}{1+s} \right\}\\
\cong &\max_{1\ge s \ge 0} \left\{ 
\frac{-I(p;W)s - J(p;W)s^2+(I(p;W)+\Delta)s }{1+s}\right\} \\
\cong &\max_{1\ge s \ge 0}\left\{
- J(p;W)s^2 + \Delta s \right\}\cong 
\frac{\Delta^2}{4 J(p;W)} \\
& \max_{0 \ge t \ge -1/2} \left\{- \phi(t|W,p) - tR \right\}\\
\cong & \max_{0 \ge t \ge -1/2} \left\{
I(p;W)t - J(p;W)t^2 - (I(p;W)+\Delta)t \right\}\\
= & \max_{0 \ge t \ge -1/2}\left\{- J(p;W)t^2 - \Delta t \right\}
= \frac{\Delta^2}{4 J(p;W)}\\
& \max_{s \ge 0}\left\{
\frac{-\psi(s|W)+sR}{1+s}\right\} \\
\cong &
\max_{s \ge 0}\Biggl\{
 \frac{-I(p_0;W) s - (J(p_0;W)+\rE_{p_0} H(W_x)) s^2}{1+s}\\
&\qquad +\frac{s(I(p_0;W)+\Delta)}{1+s} \Biggr\}\\
\cong &
\max_{s \ge 0}\left\{
- (J(p_0;W)+\rE_{p_0} H(W_x)) s^2 + \Delta s \right\}\\
= &\frac{\Delta^2}{4 (J(p_0;W)+\rE_{p_0} H(W_x))}.
\end{align*}
\end{proof}

\section{Conclusion}
We give several non-asymptotic formulas
in identification code, channel resolvability, and
wire-tap channel.
Using these formulas,
we give the achievable rate channel resolvability
for the general channel, which had been an open problem.
Also, we derived several asymptotic relations among
divergence rates, capacities of identification code,
and $\epsilon$ capacities of channel resolvability.

From these non-asymptotic formulas,
we obtained lower bounds of error exponents of 
channel resolvability in the stationary memoryless setting.
Moreover, we derived lower bounds of 
error probability and wire-tapper's information
in the stationary memoryless setting in
wire-tap channel.

Concerning the quantum setting,
wire-tap channel has been discussed in the discrete memoryless channel case
by Devetak \cite{Deve}, Winter {\it et. al.}\cite{WNI}
and Cai \& Yeung\cite{CY},
and identification codes has been discussed by Ahlswede \& Winter\cite{AW}.
Hence, several quantum extensions of the results presented here
can be expected.
Some has been obtained by the author.
And some of them have appeared in the author's textbook\cite{H-text}.
Those not already presented 
will appear in a forthcoming paper.

\section*{Acknowledgments}
The author would like to thank Professor Hiroshi Imai of the
QCI project for support.
He is grateful to Professor Yasutada Oohama for 
useful discussions.
He is also grateful to reviewers
for their kind and useful comments.

\bibliographystyle{IEEE}

\begin{biography}{Masahito Hayashi} was born in Japan in 1971.
He received the B. S. degree from Faculty of Sciences in Kyoto 
University, Japan, in 1994
and the M. S. and Ph. D. degrees in Mathematics from 
Kyoto University, Japan, in 1996 and 1999, respectively.

He worked in Kyoto University as a Research Fellow of the Japan Society of the 
Promotion of Science from 1998 to 2000,
and worked in the 
Laboratory for Mathematical Neuroscience, 
Brain Science Institute, RIKEN from 2000 to 2003.
In 2003, he joined 
Quantum Computation and Information Project, ERATO, JST
as the Research Head.
He also works in 
Superrobust Computation Project
Information Science and Technology Strategic Core (21st Century COE by MEXT)
Graduate School of Information Science and Technology
The University of Tokyo
as Adjunct Associate Professor from 2004.
He is an Editorial Board of International Journal of Quantum Information.
His research interests include quantum information theory and
quantum statistical inference.
\end{biography}
\end{document}